\documentclass[12pt]{article}
\usepackage[utf8]{inputenc}
\usepackage{mathrsfs}
\usepackage{changepage}
\usepackage{multirow}
\usepackage{amsmath}
\usepackage{array}
\usepackage{booktabs}
\usepackage{threeparttable}
\usepackage{rotating}
\usepackage{natbib}
\usepackage{multicol}
\usepackage{graphicx}
\usepackage{tipa}
\usepackage{amssymb}
\usepackage{bm}
\usepackage[english]{babel}
\usepackage{amsthm}
\usepackage{algorithm}
\usepackage{algorithmic}
\usepackage{tikz}
\usetikzlibrary{shapes,arrows}
\usepackage{setspace}
\usepackage{comment}

\def\logit{\mathrm{logit}}
\def\diag{\mathrm{diag}}
\def\exp{\mathrm{exp}}

\usepackage[title]{appendix}

\newcommand{\blind}{0}

\begin{document}

\def\spacingset#1{\renewcommand{\baselinestretch}%
{#1}\small\normalsize} \spacingset{1}

\if0\blind
{
  \title{\bf Stochastic Volatility under Informative Missingness}
  \author{Gehui Zhang \\
    Southwest Petroleum University \& \\
    Department of Biostatistics, University of Pittsburgh \\ 
    Gong Tang \\
    Department of Biostatistics, University of Pittsburgh\\
    Lori N. Scott \\
    Department of Psychiatry, University of Pittsburgh\\
    and\\
    Robert T. Krafty\thanks{
    Corresponding author Robert T. Krafty, Department of Biostatistics and Bioinformatics, Emory University, Atlanta, GA 30322 (e-mail:rkrafty@emory.edu). This work is supported by National Institutes of Health grants R01MH115388, R01GM140476 and R01GM113243.} \\
    Department of Biostatistics and Bioinformatics, Emory University}
  \maketitle
  \thispagestyle{empty}

} \fi

\if1\blind
{
  \bigskip
  \bigskip
  \bigskip
  \begin{center}
    {\LARGE\bf Stochastic Volatility under Informative Missingness}
\end{center}
  \medskip
  \thispagestyle{empty}

} \fi
%\clearpage
%\pagenumbering{arabic} 
\bigskip
%\newpage
\begin{abstract}
Stochastic volatility models that treat the variance of a time series as a stochastic process have proven to be important tools for analyzing dynamic variability.  Current methods for fitting and conducting inference on stochastic volatility models are limited by the assumption that any missing data are missing at random. With a recent explosion in technology to facilitate the collection of dynamic self-response data for which mechanisms underlying missing data are inherently scientifically informative, this limitation in statistical methodology also limits scientific advancement. The goal of this article is to develop the first statistical methodology for modeling, fitting, and conducting inference on stochastic volatility with data that are missing not at random. The approach is based upon a novel imputation method derived using Tukey's representation, which utilizes the Markovian nature of stochastic volatility models to overcome unidentifiable components often faced when modeling informative missingness in other settings. This imputation method is combined with a new conditional particle filtering with ancestor sampling procedure that accounts for variability in imputation to formulate a complete particle Gibbs sampling scheme. The use of the method is illustrated through the analysis of mobile phone self-reported mood from individuals being monitored after unsuccessful suicide attempts.    
\end{abstract}
%\begin{center}
%\section*{Abstract}
%\end{center}

\noindent%
\vfill
%{\it KEY WORDS:}

\newpage
\spacingset{1.9} % DON'T change the spacing!
\section{Introduction}

Stochastic volatility models are nonlinear state space models where the variance or dispersion of a time series is modeled as a dynamic stochastic process.  Given the presence and importance of heteroscedastic variability in derivative options, stochastic volatility models have become a popular tool for analyzing financial and econometric time series, and a sizable amount of research has been conducted on methods for their analysis. A comprehensive description of the development and use of stochastic volatility models can be found in the first chapter of \cite{Shepard2005}.

\begin{figure}[!tb]
    \centering
    \includegraphics[width=.8\textwidth]{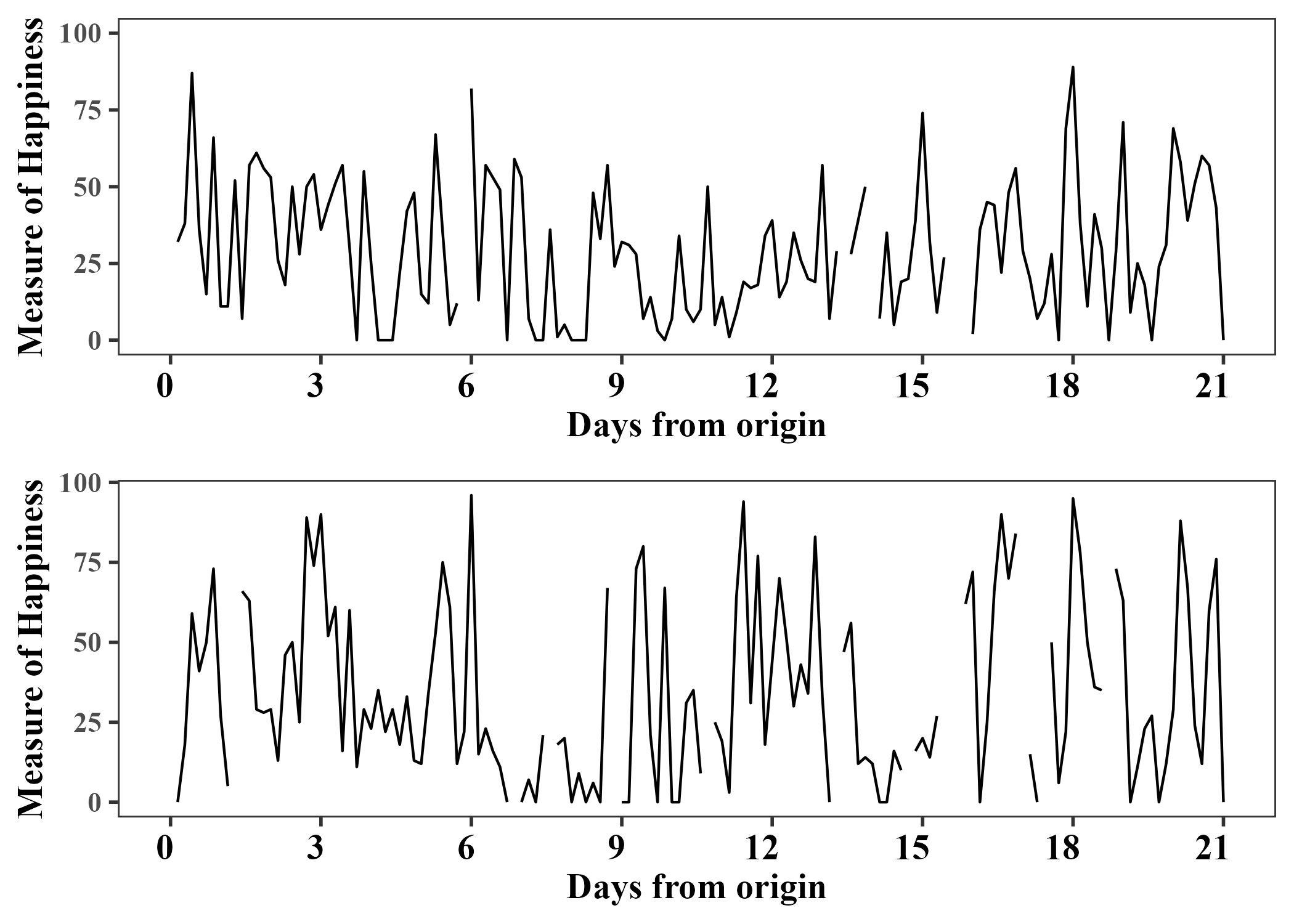}
    \caption{Self-reported happiness over time from participants who recently had suicidal ideation and/or behavior. The participants were prompted to report happiness 7 times per day over 3 weeks. One hundred (100) represents maximum happiness, and 0 represents minimum happiness.}
    %released from an in-patient facility after an unsuccessful suicide attempt
    \label{fig:happyori}
\end{figure}

Technological advances have led to an explosion in the amount of dynamic data that are collected across a variety of fields. An important example of such data is activity and emotional data collected by mobile devices in real time, which are often referred to as ecological momentary assessment (EMA) data. Researchers have found associations between variability in mood and affect with clinical outcomes \citep{Renee11, Kuppens, McConville, Angst}.  However, the current state of the art for evaluating variability in psychosocial and behavioral data either assumes homoscedastic variability, despite empirical evidence of heteroscedasticity, or utilizes coarse and naive windowed estimates within predefined time intervals. Stochastic volatility could be a potentially interpretable and predictive metric from EMA data that provides a more comprehensive, dynamic picture of variably. An example of such data comes from our motivating application, which is discussed in further detail in Section \ref{application}, which considers data from two young adults who experienced recent suicidal thoughts and/or behaviors \citep{DEAR23}. The participants were prompted to self-report measures of affect and mood seven times per day using a cell phone-based application. The time series of the participants' self-reported level of happiness are displayed in Figure \ref{fig:happyori}. We desire an analysis of these data that can provide estimates of stochastic emotional volatility, which can potentially be used to predict clinical outcomes, such as suicide ideation. 

The observational nature of self-reported data prompted by mobile devices inherently results in nonignorable missingness. However, existing methods for estimating stochastic volatility assume any missing data are missing at random. For financial time series where the ignorability of missing data is plausible, $ad ~ hoc$ methods that assume data are missing completely at random, such as ignoring missing points, averaging, or aggregation, are commonly used to handle missing data. A technique for handling irregular sampling in data that uses the expectation-maximization (EM) algorithm within particle Gibbs sampling was proposed by \cite{KimJ08}, which can be used to analyze times series with missing data by assuming that missing observations do not relate to latent volatility and are missing at random. To the best of our knowledge, no methods currently exist for stochastic volatility models with missing data that are not missing at random. The broad goal of this article is to develop the first formal method for the analysis of stochastic volatility with informative missing data. 
%We propose a method to handle informative missingness using Tukey's representation and a modified particle Markov chain Monte Carlo method utilizing the conditional particle filter with ancestor sampling. 

Selection factorization \citep{Rubinselect} and pattern-mixture factorization \citep{Rubinselect, Littlepattern} are arguably the two most popular approaches for addressing nonignorable missingness. However, both of these approaches involve an unidentifiable component that has to come from prior knowledge or remain unknown. \cite{Holland} demonstrated that this unidentifiability is propagated into other components of these two popular factorizations. To address this challenge, as recorded by \cite{Holland}, John W. Tukey proposed an alternative method for decomposing the joint distribution in which %Tukey's representation or simplified selection model(SS) only needs to specify the missing mechanism and the distribution of observed data.
all components of the representation are accessible. Tukey's representation was recently re-examined by \cite{Frank20}, who demonstrated its tractability when data are independent (i.e. not time series) and the distribution of the variable of interest is in the exponential family. One of two specific methodological contributions of our article is the development of a novel imputation procedure for the stochastic volatility model based on Tukey's representation. Although it provides a factorization with identifiable components, a potential challenge in the use of Tukey's representation is the need to estimate the conditional probability of missingness utilizing all data, both observed and unobserved. We show that the Markovian state space nature of the stochastic volatility model allows one to formulate and estimate this quantity using either a parametric or nonparametric logistic regression model.  To the best of our knowledge, all past usages of the Tukey's representation have only considered parametric models for the missingness mechanism that are dictated by the natural parameter of the exponential family considered. Our method not only represents what we believe is the first use of Tukey's representation for time series, but also its first usage with nonparametric modeling of the missing-data mechanism.

As a nonlinear state space model, there are a variety of techniques to approach Bayesian analysis for the stochastic volatility model. \cite{Carlin92}, \cite{Eric94}, \cite{Taylor94}, and \cite{KimS98} discussed classic Markov chain Monte Carlo (MCMC) methods for the stochastic volatility model. Particle Gibbs methods, also commonly known as sequential Monte Carlo, have become a popular approach to fitting stochastic volatility models. \cite{Andrieu10} proposed using the conditional particle filter (CPF) for the latent state, which guarantees that the target distribution is invariant, unlike previous methods. Later, \cite{Lindsten14} proposed the conditional particle filter with ancestor sampling (CPF-AS) to overcome the path degeneracy common with CPF. The second specific methodological contribution of this article is the development of an imputed CPF-AS algorithm (ICPF-AS) that incorporates information from the missingness mechanism and presents a complete particle Gibbs method for the analysis of stochastic volatility models with informed missingness.  To the best of our knowledge, the novel ICPF-AS is the first method for fitting stochastic volatility models with informed missingness. It should be noted that, as discussed in Section \ref{discussion}, although formulated for the analysis of the stochastic volatility model, the proposed method provides insights into the analysis of general nonlinear state space models with informed missingness. 

The rest of this article is presented as follows.
After introducing the univariate discrete time stochastic volatility model with missing data in Section \ref{model}, a likelihood decomposition is derived in Section \ref{PG_Miss} to facilitate and motivate the proposed methodology.  The imputation method is constructed in Section \ref{impy}, the new ICPF-AS to account for imputation is introduced in  Section \ref{icpf}, while priors distributions and the complete particle Gibbs sampling scheme are discussed in Section \ref{complete}. Simulation studies are presented in Section \ref{simulation} to investigate the approach's empirical properties, and its use in the analysis of EMA data is illustrated in Section \ref{application}. A discussion of extensions, limitations, and further implications is provided in Section \ref{discussion}. Appendices provided in online supplemental materials provide details of the Gibbs sampler and additional simulation results.

\section{Model}\label{model}

The basic univariate discrete time stochastic volatility model for a real-valued zero-mean time series $\{y_t\}_{t=1}^n$ defined as
\vspace{-0.3\baselineskip}
\begin{equation}  \label{usvm}
    \begin{aligned}
        y_t & =\exp\left(\frac{h_t+\mu}{2}\right) \epsilon_t \\
        h_{t+1} & =\mu+\phi \left(h_{t}-\mu\right)+\eta_t       
    \end{aligned}
\end{equation}
where $\epsilon_t \sim N(0,1)$, $\eta_t \sim N(0,\sigma^2)$, and $h_1 \sim N\left(\mu,\frac{\sigma^2}{1-\phi^2}\right)$, $\epsilon_t,\,\eta_t,\,j_1$ are independent from each other. Here, $\left\{h_t\right\}_{t=1}^n$ is the volatility process and the primary focus of our analysis. In this article, we will utilize the notation where $h_{1:n}$ and $y_{1:n}$ represent the epochs of the stochastic processes $\{h_t\}_{t=1}^n$ and $\{y_t\}_{t=1}^n$. 

\cite{shephard96} and \cite{Taylor94} discussed theoretical details and econometric properties of this model when $y_t$ are observed for all $t=1,\dots,n$. This article considers the more complicated setting where some values of $y_{1:n}$ could be missing. Define the response indicator at time $t$, $r_t$, as:
\[ r_t =
  \begin{cases}
    1       & \quad \text{if } y_{t} \text{ is observed}\\
    0  & \quad \text{if } y_{t} \text{ is missing}.
  \end{cases}
\]
We assume that $y_1$ is always observed so that $r_1=1$. We are interested in the scenario where the probability of missing data at time $t$ only depends on $y_t$, so that the presence of missing data is informative. We assume that $r_t$ only depends on $h_{1:n}$ and $y_{1:n}$ through $y_t$, and that $\left\{r_t \right\}_{t=1}^n$ are independent conditional on $y_{1:n}$. 

We consider that the basic stochastic volatility model \eqref{usvm} holds for the observed $y_t$ given $r_t=1$, and believe the conditional distribution is different when $y_t$ is missing. Further, we consider the scenario where a model for the conditional log odds of $r_t = 1$ based on $y_t$ can be assumed such that
\begin{equation}\label{logitassump}
    \logit\left[f\left(r_t=1 \mid y_{1:n},\theta_{R|Y}\right)\right]=g\left(y_t \mid \theta_{R|Y}\right)
\end{equation}
for some function $g$ of $y_t$ that depends on a set of parameters $\theta_{R|Y}$. We consider two modeling approaches for \eqref{logitassump} in this article. 
The first approach is the simple parametric case where $\theta_{R|Y} = \left(\beta_0, \beta_1\right)^T$ and $g\left(y_t |{\theta_{R|Y}}\right) = \beta_0 + \beta_1 y_t$, which accurately captures the hypothesized mechanism of the motivating application where lower levels of positive affect are more likely to be missing. As will be demonstrated in the next section, this setting provides for a simple, closed form for the distribution of $y_t$ when $r_t=0$.  

The second approach is a nonparametric approach that models $g$ as a smooth function with square integrable second derivatives through the Bayesian formulation of the cubic smoothing spline.  Here, we model $g\left(y_t |{\theta_{R|Y}}\right) = \beta_0 + \beta_1 y_t + u(y_t)$, where $u$ is a function with square-integrable second derivatives that is orthogonal to a space of linear functions in that $\int u(y) dy = \int u'(y) dy = 0$, and
$\theta_{R\mid Y} = \left(\beta_0, \beta_1, u \right)$.  This model decomposes $g$ into a linear component and a component that accounts for curvature. Under our fully Bayesian approach that will be defined in detail in Section \ref{complete}, weakly informative Gaussian priors are assumed for $\beta_0$ and $\beta_1$, and $u$ is modeled as a mean-zero Gaussian process whose covariance function (after scaling its domain between [0,1]) is $\lambda^{-1} R\left(x, y\right)$ where $R\left(x,y \right) = k_2(x)k_2(y) - k_4(|x - y|)$, $k_2(x)=\frac{1}{2}\left[ (x-0.5)^2-\frac{1}{12}\right]$ and $k_4(x)=\frac{1}{24}\left[ (x-0.5)^4-\frac{(x-0.5)^2}{2}+\frac{7}{240}\right]$ \citep[Chapter 2]{SS:ChongGuSS}. The smoothing parameter $\lambda$ regularizes the roughness of $g$ such that, as $\lambda \rightarrow 0$, $u$ almost surely approaches a constant function of zero, and $g$ approaches a linear function.  Although any reasonable nonparametric approach could conceivably be considered, we consider the cubic smoothing spline as its decomposition into linear and nonlinear components, and the closed form under a linear model, enables for the intuitive and efficient formulation of an important sampler.

\section{Implied distribution for missing data}\label{impy}

%In order to sample from $q(x_t^0,h_t\mid h_{t-1}, \Theta)$, we need to figure out the implied distribution of missing data $f(y_t\mid r_t=0, h_{t-1}, \Theta)$. 
With a slight abuse of notation that makes the dependence on $h_t$ and $\Theta$ implicit, let
\begin{align*}
    % f_1(y_t) &= f(Y_t=y_t|R_t=1,\theta_{Y|R}, h_{t-1})\\
    % f_0(y_t) &= f(Y_t=y_t|R_t=0,\theta_{Y|R}, h_{t-1})\\
    % r(y_t) &= P(R_t=1|Y_t=y_t,\theta_{R|Y}, h_{t-1})\\
    % p &= P(R_t=1|\theta_R)
    r(y_t) &= p(r_t=1|y_t,\Theta, h_{t})\\
    p &= p(r_t=1|\Theta, h_{t}).
\end{align*}
 If the positivity condition is satisfied where $P(R_t=r_t|\Theta)>0$ and $P(Y_t=y_t|\Theta)>0$ such that $P(R_t=r_t,Y_t=y_t|\Theta)>0$ \citep{Holland}, then Tukey's representation expresses the conditional distribution of missing $y$ as
\begin{equation} \label{tukeysvm}
   f(y_t|r_t=0,\Theta, h_{t})=\frac{p}{1-p}\frac{1-r(y_t)}{r(y_t)}f(y_t|r_t=1,\Theta, h_{t})
\end{equation}
(details provided in the supplementary materials). The term $\frac{p}{1-p}$ is a normalizing constant that can be determined by normalizing the integral of $f(y_t|r_t=0,\Theta, h_{t})$ to unity. Thus, the unidentifiable $f(y_t|r_t=0,\Theta, h_{t})$ is expressed in terms of specified elements that can be estimated from observed data. 

To understand each of these specified elements, we begin by expressing the density of $y_t$ given $(\Theta, h_{t}, \bm{Y}^1)$ using the total probability rule as
\vspace{-0.3\baselineskip}
\begin{equation*}
    \begin{aligned}
        f(y_t|\Theta, h_{t}, \bm{Y}^1) 
        %& = f(y_t|r_t=1,\Theta, h_{t}, \bm{Y}^1)f(r_t=1|\Theta, h_{t})\\
        %& +f(y_t|r_t=0,\Theta, h_{t}, \bm{Y}^1)f(r_t=0|\Theta, h_{t})\\
        & = pf(y_t|r_t=1,\Theta, h_{t})\\
        &+ (1-p)\left[\frac{p}{1-p}\frac{1-r(y_t)}{r(y_t)}f(y_t|r_t=1,\Theta, h_{t})\right].
    \end{aligned}
\end{equation*}
Integrating over $y_t$, it can be found that $(1-p)/p = \int f(y_t|r_t=1,\Theta, h_{t}) \frac{1-r(y_t)}{r(y_t)}\mathrm{d}y_t$.
\begin{comment}
\begin{equation*}
    p=\left[1+\int f(y_t|r_t=1,\Theta, h_{t}) \frac{1-r(y_t)}{r(y_t)} \mathrm{d}y_t\right]^{-1}
\end{equation*}
and

\begin{equation}
    \frac{p}{1-p} = \frac{1}{\int f(y_t|r_t=1,\Theta, h_{t}) \frac{1-r(y_t)}{r(y_t)}\mathrm{d}y_t}.
\end{equation}
\end{comment}
In light of our assumption that the basic stochastic volatility model \eqref{usvm} is applicable to the observed $y_t$ when $r_t=1,$
\begin{equation*}
        \begin{aligned}
    y_t \mid r_t=1,\Theta, h_{t} & \sim N(0,\sigma_t^2) \\
    \sigma_t^2 &=\exp\left(h_{t}+\mu\right). %\\
    %\eta_t &\sim N(0,\sigma^2),
        \end{aligned}
\end{equation*}
%where $\sigma_t^2 =\exp\left[\phi h_{t-1}+(1-\phi)\mu+\eta_t\right]$, $\eta_t \sim N(0,\sigma^2)$.
Under the simple logistic-linear regression model where $\theta_{R | Y} = \left(\beta_0, \beta_1\right)'$ and $g\left(y_t \mid \theta_{R | Y}\right) = \beta_0 + \beta_1 y_t$ in Equation \ref{logitassump},  $p/\left(1-p\right) =\exp \left(\beta_0-\frac{\beta_1^2\sigma_t^2}{2}\right)$.
\begin{comment}
\begin{equation} \label{normalcons}
    \frac{p}{1-p} =\exp \left(\beta_0-\frac{\beta_1^2\sigma_t^2}{2}\right).
\end{equation}
\end{comment}
Consequently, it follows that
%sing \eqref{tukeysvm} and \eqref{normalcons}, the implied distribution for missing $y_t$ is 
\begin{equation*}\label{impydens}
    f(y_t|r_t=0,\Theta, h_{t}, \sigma^2_t)  = \frac{1}{\sqrt{2\pi \sigma_t^2}}\exp \left[-\frac{(y_t+\beta_1 \sigma_t^2)^2}{2\sigma_t^2}\right] \sim N(-\beta_1\sigma_t^2,\sigma_t^2),
\end{equation*}

As expected, the implied distribution depends on the association of the probability of missingness, with imputed values having an expected value of zero when there is no association with $\beta_1 = 0$, a positive expected value when the odds of missingness is larger for larger values of $y$ with $\beta < 0$, and a negative expected value when the odds of missingness are larger for smaller values of $y$ with $\beta > 0$. Both the variance and the magnitude of the mean of the imputation distribution are proportional to $\sigma^2_t$, whose expected magnitude depends on the current time's volatility $h_{t}$.
%as well as the strength of association of the current volatility on past volatility, which is quantified through the autoregressive parameter $\phi$.  
Large positive values of current volatility,
%and positive autocorrelation
as well as large negative values of current volatility,
%and negative autocorrelation
imply larger expected values of $\sigma_t^2$. 
%As a result, the complete data distribution is a mixture of normal distribution,
%\begin{equation}
%    y_t \mid \Theta, h_{t}, \sigma^2_t \sim p\times N(0, \sigma^2_t) + (1-p)\times N(-\beta_1\sigma_t^2,\sigma_t^2)
%\end{equation}

Under the cubic smoothing spline approach where $g\left(y_t |{\theta_{R|Y}}\right) = \beta_0 + \beta_1 y_t + u(y_t)$, 
\begin{equation}
\label{SS:impliedy}
    f(y_t|r_t=0,\Theta, h_{t})
    \propto\exp\left[-u(y_t)\right] \times N(-\beta_1 \sigma_t^2,\sigma_t^2).
\end{equation} 
Note that the nonlinear structure in the cubic smoothing spline alters the implied distribution, thereby affecting the estimation procedure for both volatility and imputation steps. Whereas the distribution under missingness for the parametric log-linear model is Gaussian from which observations can be directly simulated, this is not the case under the nonparametric model.  However, as the imputation procedure is embedded into a larger sampler that utilizes importance sampling, this does not cause any complications in its use. 

\section{Particle Gibbs for SVM with missingness}

\subsection{Particle Gibbs for SVM}\label{PG_Comp}
Particle filter (PF) methods \citep{pitt1999_PF, doucet2011_PF}, also known as sequential Monte Carlo methods, are a class of recursive algorithms used for filtering, smoothing, and estimating states in nonlinear and non-Gaussian dynamic systems. These methods approximate the posterior distribution using a set of $N$ weighted random samples (particles), which are iteratively propagated through time based on a sequence of observations to provide estimates of the underlying state.
PF methods blend sequential importance sampling (SIS) and resampling \citep{Andrieu10}. In the SIS step, $N$ particles are drawn from the proposal density and assigned important weights that account for the discrepancy between the true density and proposal density. In the resampling step, a new set of $N$ particles is generated by sampling $N$ times from the particles obtained in the SIS step.

Conditional particle filter (CPF) methods, also referred to as Particle Gibbs methods, are similar to the standard PF methods but with the difference that a reference trajectory is specified and retained throughout the sampling procedure \citep{Andrieu10}. To address the poor mixing problem due to path degeneracy for CPF methods, \cite{Lindsten14} proposed conditional particle with ancestor sampling  (CPF-AS) methods. In CPF-AS methods, an artificial history is assigned to the last particle in an ancestor sampling step. 

In the absence of any missing data, Bayesian approaches to analyzing the stochastic volatility model conduct inference based on simulated realizations from conditional posterior distributions through a Gibbs sampler that iterates between sampling $\theta_{Y|R} = \left(\mu, \sigma^2, \phi\right)$ given $(h_{1:n},y_{1:n})$ and sampling $h_{1:n}$ given $\left(\theta_{Y|R}, y_{1:n} \right)$. The particle methods can be utilized for sampling sequentially from the target distribution $f\left(h_{1:n} \mid y_{1:n}, \theta_{Y|R} \right)$ utilizing a proposal distribution $q\left(h_t \mid h_{t-1}, y_t, \theta_{Y|R}\right)$ based on the  decomposition 
\begin{align}
    f(h_{1:n}\mid y_{1:n}, \theta_{Y|R}) %&=\frac{f(h_{1:n},y_{1:n}\mid \theta_{Y|R})}{f(y_{1:n}\mid \theta_{Y|R})} \nonumber\\
    & \propto f(h_{1:n},y_{1:n}\mid \theta_{Y|R})\nonumber\\
    %&=f(h_{1:n-1},y_{1:n-1}\mid \theta_{Y|R})f(h_n,y_n\mid h_{1:n-1},y_{1:n-1}, \theta_{Y|R})\nonumber\\
    %&=f(h_{1:n-1},y_{1:n-1}\mid \theta_{Y|R})f(h_n\mid h_{n-1}, \theta_{Y|R})f(y_n\mid h_n, \theta_{Y|R})\nonumber\\
    &=f(h_{1:n-1},y_{1:n-1}\mid \theta_{Y|R})\frac{f(h_n\mid h_{n-1}, \theta_{Y|R})f(y_n\mid h_n, \theta_{Y|R})}{q(h_n|h_{n-1},y_n, \theta_{Y|R})}\nonumber\\
    &\times q(h_n|h_{n-1},y_n, \theta_{Y|R}).
\end{align}
If we consider $N$ particles and let $\tilde{h}^i_{t-1}$ be the resampled-derived ancestor of $h_{t}$, details of which will be provided in Algorithm \ref{alg:CPF-AS-2}, the weight for the $i$th particle at $t$ is defined as 
\begin{equation*} 
        w_t^i= \frac{f(y_t|h_t^i,\theta_{Y|R})f(h_t^i|\tilde{h}_{t-1}^i, \theta_{Y|R})}{q(h_t^i|\tilde{h}_{t-1}^i,y_t, \theta_{Y|R})}.
\end{equation*}
Although one can choose many appropriate proposal distributions, as is common for state space models, the proposal distribution can be selected to be the state equation so that $q(h_t^i|\tilde{h}_{t-1}^i,y_t, \theta_{Y|R}) = f(h_t^i|\tilde{h}_{t-1}^i, \theta_{Y|R})$, which is $N\left[\mu + \phi \left(\tilde{h}^{i}_{t-1} - \mu \right), \sigma^2 \right]$ for $t>1$, and the weight functions simplify to $w_t^i = f\left(y_t \mid h^i_t, \theta_{Y|R} \right)$.

\begin{comment}
Denote the state formula in model \eqref{model} as: 
\begin{equation*}  \label{usvm for alg}
    %\begin{aligned}
        h_{t} \sim f_{\theta}(h_{t}|h_{t-1}) = N \left[ \mu + \phi \left(h_{t-1} - \mu \right), \sigma^2 \right], \,t=2,\dots,n
        %y_t & \sim g_{\theta}(y_t|h_t)
    %\end{aligned}
\end{equation*}
where $\theta=(\mu,\phi,\sigma^2)$, and $h_1 \sim \pi_{\theta}(h_1) = N \left[ \mu, \sigma^2 / \left(1 -\phi^2 \right)\right]$. Denote the proposal density for $h_t$ as $r_{\theta,t}(\cdot)$ and the number of particles as $N$. 
For the state space model without missing data, the importance weights can be expressed as  
\begin{equation*} 
        w_t^i= \frac{f(y_t|h_t^i,\Theta)f_{\theta}(h_t^i|\tilde{h}_{t-1}^i)}{r_{\theta,t}(h_t^i|\tilde{h}_{t-1}^i,y_t)}.
\end{equation*}
Although one can choose many appropriate proposal distributions, as is common for state space models, the proposal distribution can be selected to be the state equation so that $r_{\theta,t}(h_t^i|\tilde{h}_{t-1}^i,y_t) = f_{\theta}(h_t^i|\tilde{h}_{t-1}^i)$, and the weight functions simplify to $w_t^i = f\left(y_t \mid h^i_t, \Theta \right)$.
\end{comment}

\subsection{Particle Gibbs for SVM with missingness}\label{PG_Miss}

Let $\bm{Y}^1=\{y_t:r_t=1,t=1...n\}$ and $\bm{Y}^0=\{y_t:r_t=0,t=1...n\}$ be the collection of observed and missing values up to time $n$ such that $y_{1:n} = \left\{ \bf{Y}^1, \bf{Y}^0\right\}$, and let $\Theta=\left(\theta_{Y|R}, \theta_{R|Y} \right)$ be the collection of parameters $\theta_{Y|R}$ of the stochastic volatility model and $\theta_{R|Y}$ of the logistic regression model for the probability of missingness. Our proposed Gibbs sampler that accounts for missingness-related information iterates between:
\begin{enumerate}
    \item[(i)] Sampling $\theta_{Y | R}$ given $\left(h_{1:n},\bm{Y}^1, \bm{Y}^0, \theta_{R|Y}\right)$.
    \item[(ii)] Sampling $\theta_{R | Y}$ given $\left(h_{1:n},\bm{Y}^1, \bm{Y}^0, \theta_{Y|R}\right)$.
    \item[(iii)] Sampling $(\bm{Y}^0, h_{1:n})$ given $\left(\bm{Y}^1, \Theta\right)$.   
\end{enumerate}
Step (i) can be successfully achieved through existing methods that have been used for estimating static parameters in the stochastic volatility model without missing data, and it follows from Equation \ref{logitassump} that step (ii) can be achieved through logistic regression. A challenging and innovative component to solving this novel problem is in step (iii). 

We propose to sample $(\bm{Y}^0, h_{1:n})$ given $\left(\bm{Y}^1, \Theta\right)$ sequentially using particle methods that are derived in Section \ref{decomp}. 
At time $t$, 
\begin{enumerate}
\label{icpf-sample scheme}
    \item [(i)] if $y_t$ is observed,  sample $h_t$ from proposal density $f(h_t\mid h_{t-1}, \Theta)$ and calculate the importance weights $w_t=f(y_t\mid h_t, \Theta)$, and
    \item [(ii)] if $y_t$ is missing,  sample $(y_t, h_t)$ from proposal density $q(y_t, h_t\mid h_{t-1},\Theta)=q(y_t\mid h_t,\Theta)f(h_t\mid h_{t-1}, \Theta)$ and compute the corresponding importance weights $w_t = 1$ for the parametric logistic linear setting, and $w_t=\exp\left[-u(y_t)\right]$ for the nonparametric smoothing spline setting. 
\end{enumerate}

The weights when $r_t=1$ are the same as those for PF methods when there are no missing data and quantify the likelihood of a proposed $h$ given the observed $y$.  When $r_t=0$, under the parametric logistic linear model, we can sample both $h$ and $y$ directly from their target distributions, so that particles have equal weighting and importance sampling is trivial.  Under the nonparametric smoothing spline model, we can sample $h$ directly from its target distribution, but cannot sample $y$ directly from its target distribution; the weights reflect the differences between the proposal and target distributions of $y$.

\subsubsection{Decomposition of the target distribution}\label{decomp}
While the dimensions of the proposal density differ for the proposed sequential sampler depending on $r_t=0$, where we want to sample $h_t$, or $r_t=1$, where we want to sample $\left(y_t, h_t \right)$, a prerequisite for basic particle methods is that the target distribution has a fixed dimension.  We derive the presented sampler by embedding it into a larger problem with a fixed dimension.  Consider  $\{x_t\}_{t=1}^n$ defined as 
\vspace{-0.3\baselineskip}
\begin{align}
\label{modelx}
    x_t & \sim \exp\left[-u(x_t)\right] \times N(-\beta_1 \sigma_t^2,\sigma_t^2)
    %x_t & = -\beta_1\sigma_t^2+\sigma_t\epsilon_t\\
    %\sigma_t^2 &= \exp\left(h_{t}+\mu\right)%\\
    %\epsilon & \sim N(0,1)
    %  x_t^1=&g_{obs}(h_t,\epsilon_t) \nonumber\\
    %  x_t^0=&g_{miss}(h_t,\epsilon_t) \nonumber\\
    % h_t=&\mu+\phi(h_{t-1}-\mu)+\eta_t
\end{align}
where $\sigma_t^2 = \exp\left(h_{t}+\mu\right)$ and $u(x) = 0$ in the simple logistic-linear case. The series $\{x_t\}_{t=1}^n$ is generated under the missing distribution at all time points and is never observed.  When $r_t=0$, $y_t = x_t$; when $r_t=1$, $x_t$ is a counterfactual that could have occurred if no data were observed. Sequentially sampling from the variable dimension $(h_{1:n}, Y_{1:n}^0 \mid Y_{1:n}^1, \Theta)$ can be extended to sampling from a fixed dimension $\left(h_{1:n}, x_{1:n} \mid Y_{1:n}^1, \Theta\right)$. At time $t$, 
\vspace{-0.3\baselineskip}
\allowdisplaybreaks
\begin{align}
\label{decompx}
    f(h_{1:t}, x_{1:t} \mid Y_{1:t}^1, \Theta) %&=\frac{f(h_{1:t}, x_{1:t}^0 , Y_{1:t}^1\mid \Theta)}{f(Y_{1:t}^1\mid \Theta)} \nonumber\\
    &\propto f(h_{1:t}, x_{1:t} , Y_{1:t}^1\mid \Theta) \nonumber\\
    %&=f(h_t,x_t,Y_t^1\mid h_{1:t-1}, x_{1:t-1} , Y_{1:t-1}^1, \Theta)\times f(h_{1:t-1}, x_{1:t-1} , Y_{1:t-1}^1\mid \Theta) \nonumber\\
    %&=f(h_{1:t-1}, x_{1:t-1} , Y_{1:t-1}^1\mid \Theta) \times \left[f(h_t,x_t\mid h_{1:t-1}, x_{1:t-1}^0 , Y_{1:t-1}^1 , \Theta)\right]^{1-r_t} \nonumber\\
    %&\times \left[f(h_t,x_t,y_t\mid h_{1:t-1}, x_{1:t-1} , Y_{1:t-1}^1 , \Theta)\right]^{r_t} \nonumber\\
    %&=f(h_{1:t-1}, x_{1:t-1} , Y_{1:t-1}^1\mid \Theta) \times \left[f(h_t,x_t\mid h_{t-1}, \Theta)\right]^{1-r_t}\nonumber\\
    %& \times \left[f(h_t,x_t,y_t\mid h_{t-1}, \Theta)\right]^{r_t} \nonumber\\
    &=f(h_{1:t-1}, x_{1:t-1} , Y_{1:t-1}^1\mid \Theta) \nonumber\\
    &\times \left[\frac{f(x_t\mid h_t, \Theta)f(h_t\mid h_{t-1}, \Theta)}{q(x_t,h_t\mid h_{t-1}, \Theta)}q(x_t,h_t\mid h_{t-1}, \Theta)\right]^{1-r_t} \nonumber\\
    &\times\left[\frac{f(x_t\mid h_t,y_t, \Theta)f(y_t\mid h_t, \Theta)f(h_t\mid h_{t-1}, \Theta)}{q(x_t,h_t\mid h_{t-1}, \Theta)}q(x_t,h_t\mid h_{t-1}, \Theta)\right]^{r_t}
\end{align}
where $q(x_t,h_t\mid h_{t-1}, \Theta)$ is the proposal density for the target distribution.
%The second equation follows from the Markovian structure of the stochastic volatility model, the third equation follows from the definition of $Y_t^1$, the fourth equation follows from the independence of $h_t$ on other values of $y$ conditional on $h_{t-1}$ and $y_t$, and the independence of $y_t$ and $x_t$ on $y_{1:t-1}$ or $x_{1:t-1}$ conditional on $h_{t-1}$.
%
%The decomposition \eqref{decompx} indicates the feasibility of employing a particle method to sample from the target distribution $f(h_{1:t}, x_{1:t} \mid Y_{1:t}^1, \Theta)$. 
At time $t$, the importance weight for each particle will be adjusted based on the value of $r_t$. For the $i$th particle,
\begin{align}
\label{alg-weight}
    w_t^i= W_{\Theta,t}(h_t^i,x_t^{i},r_t) & = \left[\frac{f(x_t^{i}\mid h_t^i, \Theta)f(h_t^i\mid \tilde{h}_{t-1}^i, \Theta)}{q(x_t^{i},h_t^i\mid \tilde{h}_{t-1}^i, \Theta)}\right]^{1-r_t}\nonumber\\
    &\times \left[\frac{f(x_t^{i}\mid h_t^i,y_t, \Theta)f(y_t\mid h_t^i, \Theta)f(h_t^i\mid \tilde{h}_{t-1}^i, \Theta)}{q(x_t^{i},h_t^i\mid \tilde{h}_{t-1}^i, \Theta)}\right]^{r_t}.
\end{align}
where $\tilde{h}_{t-1}^i$ is the ancestor for $(h_t^i,x_t^{i})$. However, if we choose the proposal density as $q(x_t,h_t\mid h_{t-1}, \Theta)=f(x_t\mid h_t, \Theta)f(h_t\mid h_{t-1}, \Theta)$ at time points when $r_t=1$, the sampled $x_t$ remains unused for importance weights or the inference of $x$ and $h$ series for these timepoints. 

Under the cubic smoothing spline missing-data mechanism, the importance weights are given by
$ w_t^i= W_{\Theta,t}(h_t^i,x_t^{i},r_t) = \left\{\exp\left[-u(y_t)\right]\right\}^{1-r_t}$
where at instances when $r_t=1$, the importance weights incorporate the information from observed data, while when $r_t=0$, the weights can account for the discrepancy between the proposal and target density. 
Under the simple logistic-linear missing-data mechanism, the importance weights are reduced to $ w_t^i= W_{\Theta,t}(h_t^i,x_t^{i},r_t) = \left[1\right]^{1-r_t}
    \times \left[f(y_t\mid h_t^i, \Theta)\right]^{r_t}$
where now at instances when $r_t=0$, equal weights are assigned. This is because, in the latter case, sampling is done directly from the true density.
%where now at instances when $r_t=1$, the importance weights incorporate the information from observed data, while when $r_t=0$, equal weights are assigned. This is because, in the latter case, sampling is done directly from the true density.%, and there is a lack of additional information.

For the ancestor sampling step in the CPF-AS algorithm, the index for the last particle is sampled with probability 
\begin{equation*}
    P(J=i)=\frac{W_{t-1}^i f_{\theta}(h_t^{\prime},x_t^{\prime}|h_{t-1}^i)}{\sum_{k=1}^N W_{t-1}^k f_{\theta}(h_t^{\prime},x_t^{\prime}|h_{t-1}^k)},
\end{equation*}
where $(h_t^{\prime},x_t^{\prime})$ is the reference trajectory that is prespecified and retained for the CPF-AS algorithm. The probability function can be reduced to 
\begin{align}
\label{alg-asweight}
    P(J=i)&=\frac{W_{t-1}^i f_{\theta}(x_t^{\prime}\mid h_t^{\prime})f_{\theta}(h_t^{\prime}|h_{t-1}^i)}{\sum_{k=1}^N W_{t-1}^k f_{\theta}(x_t^{\prime}\mid h_t^{\prime})f_{\theta}(h_t^{\prime}|h_{t-1}^k)} \nonumber \\
    &=\frac{W_{t-1}^i f_{\theta}(h_t^{\prime}|h_{t-1}^i)}{\sum_{k=1}^N W_{t-1}^k f_{\theta}(h_t^{\prime}|h_{t-1}^k)}.
\end{align}
The first equation follows from the definition of $x_t$, and the second equation holds as $f_{\theta}(x_t^{\prime}\mid h_t^{\prime})$ is invariant across particles by definition of the conditional particle filter methods. 

We observe that the sampled $x_t$ is not utilized in the computation of any important weights or indices. Additionally, the proposal density suggests that the sampling of $h_t$ does not incorporate $x_t$, irrespective of whether $r_t=1$ or $r_t=0$. Therefore, we can create a sub-sequence $(Y_{1:n}^0, h_{1:n})$ from $(x_{1:n}, h_{1:n})$ in a way that excludes the unused $\{x_t: r_t=1\}$ from the sampling process, which constitutes our proposed sampling scheme introduced in \ref{icpf-sample scheme}.

\subsubsection{Imputed CPF-AS (ICPF-AS)}\label{icpf}
We refer to the CPF-AS methods under imputation as imputed CPF-AS (ICPF-AS). The sampling procedure of ICPF-AS involves two steps for sampling $h$: the initial step ($t=1$) and the updating step ($t=2\dots n$).  For the initial step in Algorithm \ref{alg:CPF-AS-1}, particles for $h_1$ are sampled from the proposal density and weights computed, while Algorithm \ref{alg:CPF-AS-2} sequentially utilizes re-sampling and ancestor sampling for times $t > 1$.

\begin{algorithm}[t]
\caption{Initialization step for ICPF-AS at t=1}
\hspace*{\algorithmicindent} \textbf{Input:} The reference trajectory ($h_{1:n}^{\prime}$, $\{y_{1:n}\}^{\prime}$), and $\Theta$. 
\hspace*{\algorithmicindent}  \\
\begin{algorithmic}[1] \label{alg:CPF-AS-1}
    \STATE Draw $h_1^i \sim q_{\theta,t}(h_1|y_1)$, for $i=1,\dotsc,N-1$.
    \STATE Set $h_1^N=h_1'$.
    \STATE For $i=1,...,N$, calculate the importance weights $w_1^i=W_{\Theta,1}(h_1^i,r_1)$, and get normalized weights $W_1^i$.
    \RETURN $\{ h_{1}^i \}_{i=1}^N$, $\{ W_{1}^i \}_{i=1}^N$.
\end{algorithmic}
\end{algorithm}

\begin{algorithm}[t]
\caption{Updating ICPF-AS at time $t>1$}
\hspace*{\algorithmicindent} \textbf{Input:} The reference trajectory ($h_{1:n}^{\prime}$, $\{y_{1:n}\}^{\prime}$), $\{ (h_{1:t-1}, Y^0_{1:t-1})^i \}_{i=1}^N$,  $\{ W_{t-1}^i \}_{i=1}^N$, and $\Theta$. 
\hspace*{\algorithmicindent}  \\
\begin{algorithmic}[1] \label{alg:CPF-AS-2}
    
    %\FOR{$t=2$ to $n$}
        \STATE Draw with replacement from $\{ (h_{1:t-1}, Y^0_{1:t-1})^i \}_{i=1}^N$ with probability $\{ W_{t-1}^i \}_{i=1}^N$ for $N-1$ times to generate $\{ (\tilde{h}_{1:t-1}, \tilde{Y}_{1:t-1}^0)^i \}_{i=1}^{N-1}$.
        \STATE Set $(\tilde{h}_{1:t-1}, \tilde{Y}_{1:t-1}^0)^N=(h_{1:t-1}, Y_{1:t-1}^0)^J$, where $J$ is drawn from $1,...,N$ with probability $P(J=i)=\frac{W_{t-1}^i f_{\theta}(h_t^{\prime}|h_{t-1}^i)}{\sum_{k=1}^N W_{t-1}^k f_{\theta}(h_t^{\prime}|h_{t-1}^k)}$.
        \IF{$r_t=0$}
            \STATE Sample $(h_t,y_t^0)^i \sim q(y_t^0,h_t\mid h_{t-1}, \Theta)$ for $i=1,...,N-1$, and set $(h_t,y_t^0)^N=(\tilde{h}_{t-1}, \tilde{y}_{t-1}^0)^{N}$.
        \ELSIF{$r_t=1$}
            \STATE Sample $h_t^i \sim q(h_t\mid h_{t-1}, \Theta)$ for $i=1,...,N-1$, and set $h_t^N=\tilde{h}_{t-1}^{N}$.
        \ENDIF
        \STATE For $i=1,...,N$, compute the importance weights $w_t^i= W_{\Theta,t}(h_t^i,y_t^{0i}, r_t)$ and calculate normalized weights $W_t^i=w_t^i/\sum_{j=1}^N {w_t^j}$.
    % \ENDFOR
    % \STATE Sample $K$ from $1,...,N$ with $P(K=i)=W_n^i$
    \RETURN $\{ (h_{1:t}, Y^0_{1:t})^i \}_{i=1}^N$, $\{ W_{t}^i \}_{i=1}^N$.
\end{algorithmic}
\end{algorithm}

\section{Prior Distributions and Sampling Scheme}
\label{complete}

The approaches discussed in Section \ref{PG_Miss} for sampling $\left(\bm{Y^0}, h_{1:n} \right)$, which is one component of the Gibbs sampler that iterates between sampling $\theta_{Y \mid R}$, $\theta_{R \mid Y}$ and $\left(\bm{Y^0}, h_{1:n} \right)$.  A full iteration of the Gibbs sampler, including assumed prior distributions is given by:

\begin{enumerate}
    \item[(i)] Sampling $\theta_{Y | R}$ given $\left(h_{1:n},\bm{Y}^1, \bm{Y}^0, \theta_{R|Y}\right)$

    We assume independent priors for $\mu$ and $\left( \sigma, \phi \right)$, which allows for their sequential sampling.  Assuming a diffuse prior on $\mu$ \citep{KimS98}, we sample $\mu$ from its posterior distribution
    $$\mu|\sigma^2,\phi,h_{1:n},y_{1:n} \sim N(\hat{\mu},\sigma_{\mu}^2)$$
    where $\hat{\mu}=\sigma_{\mu}^2 \left\{\frac{(1-\phi^2)}{\sigma^2}h_1+\frac{(1-\phi)}{\sigma^2}\sum_{t=1}^{n-1}(h_{t+1}-\phi h_t) \right\}$  and $\sigma_{\mu}^2=\frac{\sigma^2}{(n-1)(1-\phi)^2+(1-\phi^2)}$.

    \,\,\,\, Although individual sampling of $\sigma$ and $\phi$ has been commonly used in the past \citep{KimS98}, the mutual dependence of $\sigma$ and $\phi$ makes the use of independent priors and condition sampling inefficient. We consider an approach that samples $\sigma$ and $\phi$ jointly using a bivariate normal prior that accounts for this dependency and the Random Walk Metropolis-Hastings (RWMH) algorithm \citep{Chen21}. Details for the joint sampling of $\sigma$ and $\phi$ and a brief review of individual sampling methods are provided in the supplemental materials.
    % The first approach assumes independent priors where  
    % \begin{align*}
    % \frac{1+\phi}{2} & \sim Beta(a,b) \\
    % \sigma &\sim \text{Half-}t(v,G)
    % \end{align*}
    % for hyperparameters $a,b >\frac{1}{2}$, $v$ and $G$.  Under these priors, we iterate between sampling $\phi$ using a Metropolis-Hastings algorithm (see the supplemental materials), and sampling $\sigma^2$ from its posterior as
    % \begin{align*}
    %         \sigma^2|q_{\sigma},h_{1:n},y_{1:n},\phi,\mu & \sim \text{Inv-}\Gamma(\frac{v+n}{2},A) \\
    %  q_{\sigma}|\sigma^2,h_{1:n},y_{1:n},\phi,\mu & \sim \text{Inv-}\Gamma(\frac{v+1}{2},\frac{1}{G^2}+\frac{v}{\sigma^2})
    % \end{align*}
    % where $A=\frac{v}{q_{\sigma}}+\frac{1}{2}\{(1-\phi^2)(h_1-\mu)^2+\sum_{t=2}^n [(h_t-\mu)-\phi(h_{t-1}-\mu)^2]\}$.

    % \,\,\,\, The second approach recognizes that the mutual dependence of $\sigma$ and $\phi$ makes the use of independent priors and condition sampling inefficient.  An alternative 

    \item[(ii)] Sampling $\theta_{R | Y}$ given $\left(h_{1:n},\bm{Y}^1, \bm{Y}^0, \theta_{Y|R}\right)$.
    \begin{enumerate}
        \item Under the simple logistic missing-data mechanism, the parameters $\beta_0,\beta_1$can be sampled using the P\'{o}lya-Gamma latent variable method for Bayesian logistic regression \citep{polya}. For hyperparameters $m_0$ and $B$, we assume the prior where $\bm{\beta}\sim N(m_0,B)$. To be consistent with logistic regression, we reparameterize and consider $m_t=1-r_t, t=1,...,n$ as the outcome variable that indicates missing in the regression. Let $X= (x_1,...x_n)^{\prime}$ denote the design matrix, where $x_i=(1,y_i)$.  We then sample $\bm{\beta}$ as
        \vspace{-0.8\baselineskip}
        \begin{align*}
        z_i|\bm{\beta} & \sim PG(1,x_i^T\bm{\beta}) \\
        \bm{\beta}|m,z & \sim N(\mu_z,V_z)
        \end{align*}
        where 
        $V_z=\left(X^TZX+B^{-1}\right)^{-1}$, $\mu_z=V_z\left(X^T\kappa+B^{-1}b\right)$, $\kappa=\frac{1}{2}\textbf{1}_n$ and $Z$ is the $n \times n$ diagonal matrix with $i$th diagonal element $z_{i}$. The estimation of $\bm{\beta}=(\beta_0, \beta_1)$ is performed with standardized $y$ such that the covariance matrix $B$ can be reduced to a diagonal matrix. According to the interpretation of model parameters of a logistic regression model, we propose to choose and update the hyperparameter $m_0$ with the following procedure:
    \begin{itemize}
        \item Choose the first item of $m_0$ as $\log\left( \frac{m_{mean}}{1-m_{mean}}\right)$, where $m_{mean}$ is the mean value of $\{m_t\}_{t=1}^n$.
        \item For the second item of $m_0$, we choose an arbitrary starting value for the first iteration, and for the $i$th iteration $(i>1)$, the timepoints are split into $T^u={t:\,y_t>0.5}$ and $T^l={t:\,y_t<=0.5}$, and update the item as $\log\left( \frac{m_{mean}^u/1-m_{mean}^u}{m_{mean}^l/1-m_{mean}^l}\right)$ where $m_{mean}^u$ is the mean value of $\{m_t\}_{t\in T^u}$ and $m_{mean}^l$ is the mean value of $\{m_t\}_{t\in T^l}$.
    \end{itemize}
        \item In the context of the cubic smoothing spline missing-data mechanism, we propose to employ a computationally efficient low-rank approximation. Rescalling data between [0,1], define $y^*_t=\left[y_t-\min(y)\right]/\left[\max(y)-\min(y)\right]$, $d_1=\beta_0+\beta_1 \min(y)$, and $d_2=\beta_1\left[\max(y)-\min(y)\right]$. The right side of the missing-data mechanism can be written as 
        $g\left( \bm{Y}^*\right) =S\bm{d}+R\bm{c},$  where $ \bm{Y}^*$ is the $n$ dimensional vector with $i$th element $y^*_i$, $S$ is the $n \times 2$ matrix with $i$th row $\left(1, y_i^*\right)$, $R$ is the $n \times n$ matrix with $ij$ element $R(y^*_i, y^*_j)$, $R\left(\cdot, \cdot\right)$ is the kernel defined in Section \ref{model}, and
        $\bm{c} \sim N\left( \bm{0}, \lambda^{-1} R^{-1} \right)$.  The k-dimensional low-rank approximation considers a $k$-dimensional grid $s_j = j/k$, defines the $k \times k$ matrix $Q$ as the evaluation of the kernel $R$ over this grid, and lets $Q=U D U^T$ be its singular value decomposition.  Defining the $n \times k$ matrix $\tilde{Q}$ with $ij$th element $R(y^*_i, s_j)$ and $\tilde{R} = \tilde{Q} U^T D^{-1/2}$, the low-rank approximation considers $g\left(\bm{Y}^*\right) = S\bm{d}+\tilde{R}\tilde{\bm{c}}$ where 
        \vspace{-0.3\baselineskip}
        \begin{align*}
        (\bm{d},\bm{\tilde{c}})^T &\sim N(m_{k+2\times1},B_{(k+2)\times(k+2)}), \\
        \lambda^{-\frac{1}{2}} & \sim \text{Half-}t(\nu_{\lambda},G_{\lambda}),    
        \end{align*}
         $m_{k+2\times 1}$ is the zero vector and $\nu_{\lambda}$, $G_{\lambda}$ and $B_{(k+2)\times(k+2)} = \diag \left\{\sigma^2_d, \sigma^2_d,\lambda^{-1},\dots,\lambda^{-1} \right\}$ are hyperparameters. The half-t distribution is equivalent to a scale mixture of inverse Gamma distribution \citep{wand11}, 
        \vspace{-0.3\baselineskip}
        \begin{align*}
            \lambda^{-1}|q_{\lambda}&\sim \text{Inv-}\Gamma\left(\frac{v}{2},\frac{v}{q_{\lambda}}\right)\\
            q_{\lambda}& \sim \text{Inv-}\Gamma\left(\frac{1}{2},\frac{1}{G_{\lambda}^2}\right)
        \end{align*}
        The coefficients $\bm{d}$ and $\bm{\tilde{c}}$ are then sampled with the P\'{o}lya-Gamma latent variable method. The posterior distribution for $\lambda$ is 
        \begin{align*}
            \Omega_{\lambda}\mid \bm{d}, \bm{\tilde{c}}, \lambda&\sim \text{Inv-}\Gamma\left(\frac{\nu_{\lambda}+1}{2},\frac{1}{G^2_{\lambda}}+\nu_{\lambda}\lambda\right)\\
            %a_{n\Omega} &= \frac{\nu_{\lambda}+1}{2}\\
            %b_{n\Omega} &= \frac{1}{G^2_{\lambda}}+\nu_{\lambda}\lambda\\
            \lambda^{-1}\mid \bm{d}, \bm{\tilde{c}}, \Omega & \sim \text{Inv-}\Gamma\left(\frac{\nu_{\lambda}+k}{2},\frac{\nu_{\lambda}}{\Omega_{\lambda}}+\frac{1}{2}\bm{\tilde{c}}^T\bm{\tilde{c}}\right).
            %a_{n\lambda} &= \frac{\nu_{\lambda}+k}{2}\\
            %b_{n\lambda} &= \frac{\nu_{\lambda}}{\Omega}+\frac{1}{2}\bm{\tilde{c}}^T\bm{\tilde{c}}
        \end{align*}
    \end{enumerate}
    
    \item[(iii)] Sampling $(\bm{Y}^0, h_{1:n})$ given $\left(\bm{Y}^1, \Theta\right)$

    The parameters  $(\bm{Y}^0, h_{1:n})$ are sampled using ICPF-AS in the last part of the Gibbs sampler utilizing methods developed in Sections \ref{PG_Miss}, as formalized in Algorithm \ref{alg:MCPF-AS}.

\begin{algorithm}[t]
\caption{ICPF-AS for Sampling $\bm{Y^0}, h_{1:n}$}
\hspace*{\algorithmicindent} \textbf{Input:} The reference trajectory ($h_{1:n}^{\prime}$, $\{y_{1:n}\}^{\prime}$), and $\Theta$. 
\hspace*{\algorithmicindent}  \\
\begin{algorithmic}[1] \label{alg:MCPF-AS}
    
    \STATE Initiate the algorithm using Algorithm \ref{alg:CPF-AS-1}.
    \FOR{$t=2$ to $n$}
        %\IF{$r_t=0$}
        %    \STATE Impute $y_t$ using the Algorithm \ref{alg:impute}.
        %\ENDIF
        \STATE Update $h_t$ using the Algorithm \ref{alg:CPF-AS-2}.
    \ENDFOR
    \STATE Sample $K$ from $1,...,N$ with $P(K=i)=W_n^i$
    \RETURN $h_{1:n}^{K}$.
\end{algorithmic}
\end{algorithm}

\end{enumerate}

\section{Simulation}
\label{simulation}

\begin{figure}[!tb]
    \centering
    \includegraphics[width=0.75\textwidth]{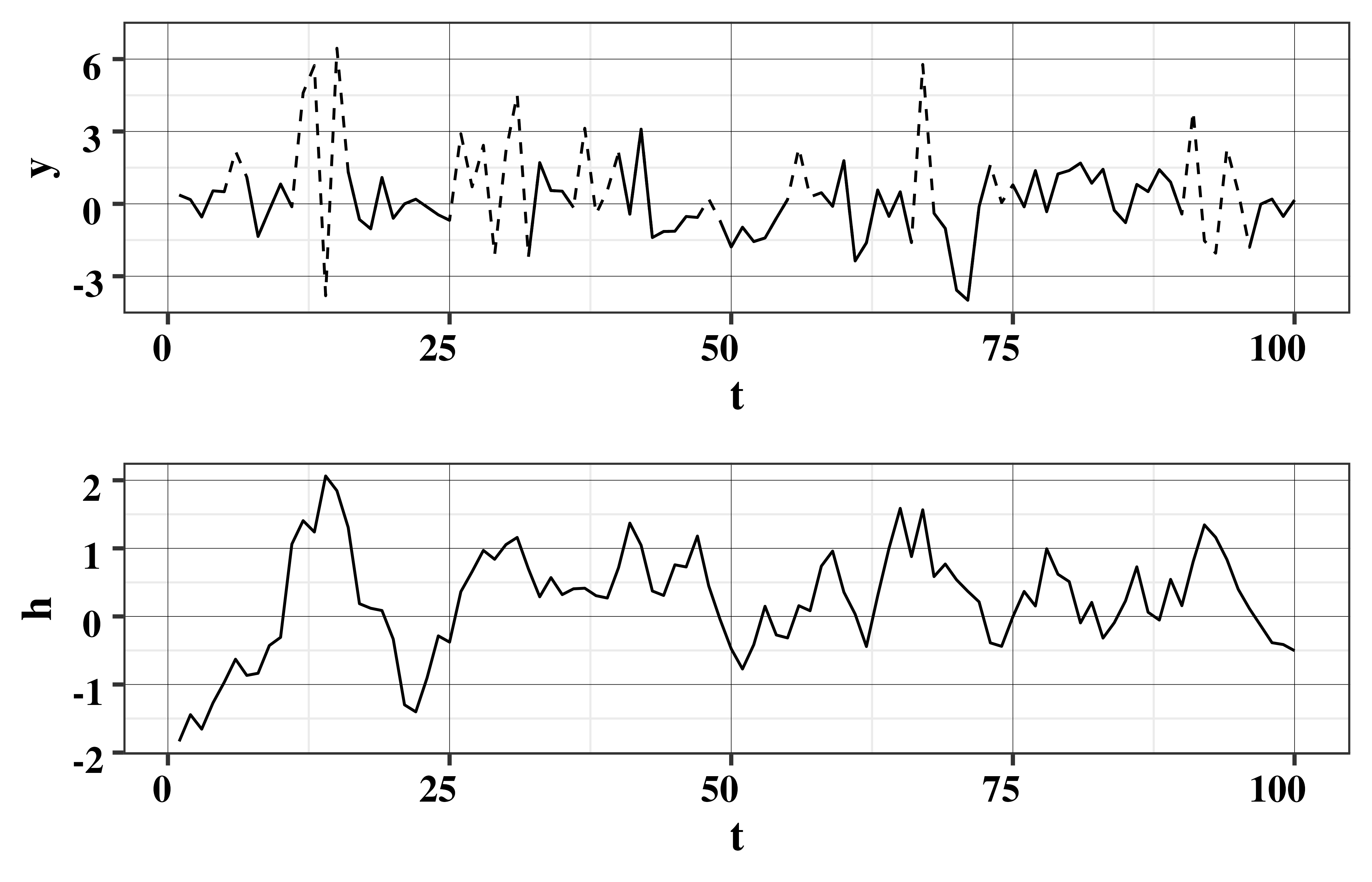}
    \caption{The volatility $h$ and time series $y$ sequence generated from model \ref{usvm} with sample size $n=100$ and $\beta_1=\ln(3)$.}
    \label{fig:sim_hy0}
\end{figure}

For the stochastic volatility model \ref{usvm}, we set the true values of $\theta_{Y\mid R}=(\mu, \sigma^2, \phi)$ as $(0.1,0.25,0.8)$ and the number of time points $n=100, 500$. For the linear missing-data mechanism, we set the parameters for the missing-data mechanism $\beta_0=-3$ and $\beta_1=\ln(2.5),\ln(3),\ln(3.5)$. For example, Figure \ref{fig:sim_hy0} shows the volatility $h$ and time series $y$ sequence generated from model \ref{usvm} with sample size $n=100$ and $\beta_1=\ln(3)$.

\begin{figure}[!tb]
    \centering
    \includegraphics[width=0.75\textwidth]{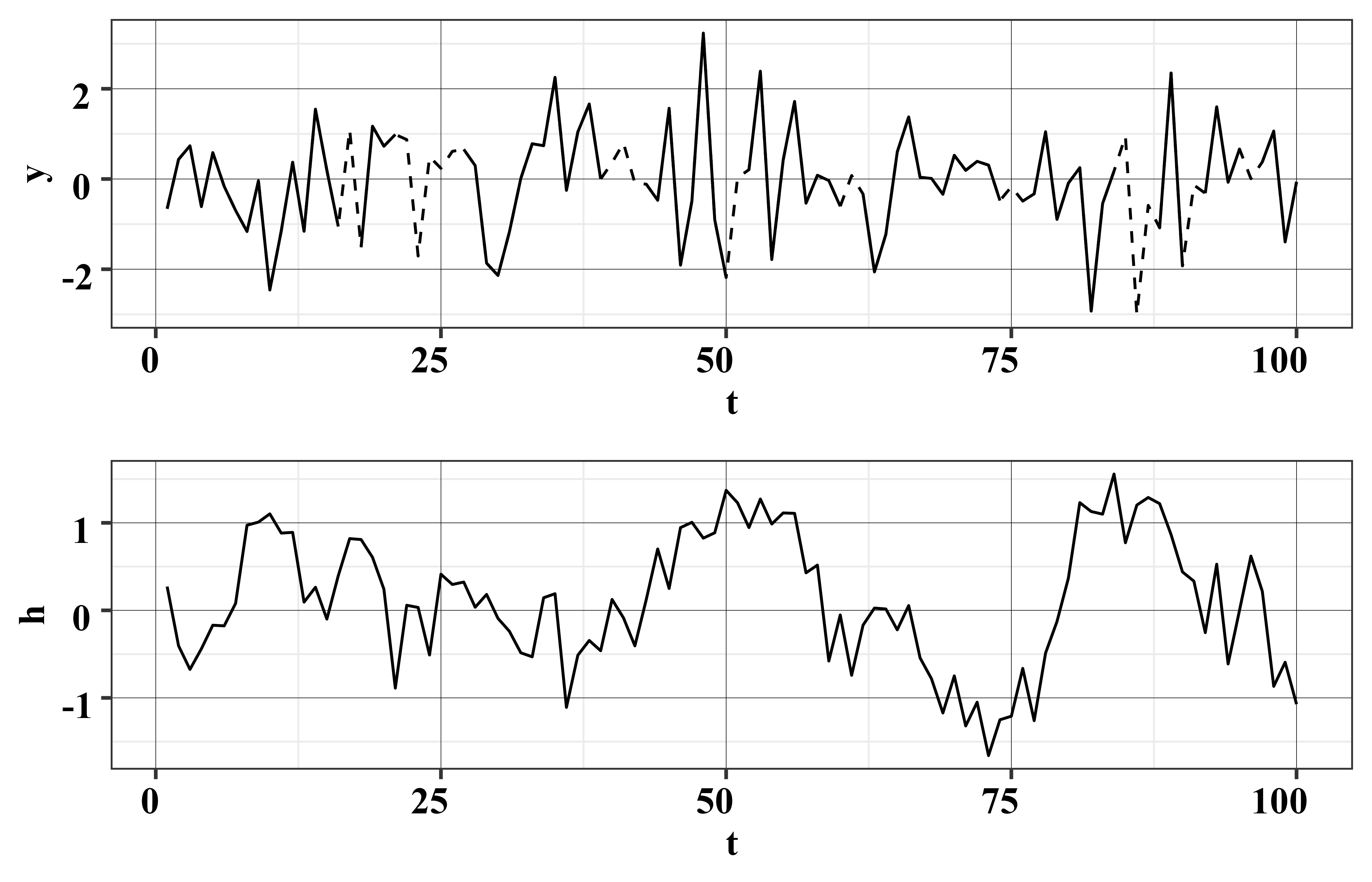}
    \caption{The volatility $h$ and time series $y$ sequence generated from model \ref{usvm} with sample size $n=100$ and smoothing spline missing-data mechanism with $\beta_1=\ln(4.5)$.}
    \label{fig:sim_hy0_ss}
\end{figure}

For the smoothing spline missing-data mechanism, the true missing-data mechanism is selected as 
\vspace{-0.8\baselineskip}
\begin{equation}
    \logit \left[ f(r_t=1 \mid y_{1:n},\theta_{R|Y})\right]=\beta_0+\beta_1y_t+y_t^2.
    \label{ss_trueeq}
\end{equation}
We set the parameters for missing-data mechanism $\beta_0=-2$ and $\beta_1=\ln(2.5),\ln(3.5),\ln(4.5)$. The number of particles is $N=20$. We run 500 replicates for each beta and sample size combination. For example, Figure \ref{fig:sim_hy0_ss} shows the volatility $h$ and time series $y$ sequence generated from model \ref{usvm} with sample size $n=100$ and $\beta_1=\ln(4.5)$.

%For sampling $\phi$ and $\sigma$ separately, we set the hyperparameters for $\phi$ and $\sigma$ as $a=20, b=1.5, v=10$, and $q_{\sigma}=0.6$. 
Regarding the hyperparameters for model parameters, we set the hyperparameters for $\phi$ and $\sigma$ as $\mu_{\phi}=0.875, \mu_{\sigma}=0.45, \sigma_{\phi}=0.075, \sigma_q=0.1$ and $\rho=-0.25$. For the linear missing-data mechanism, the first hyperparameter for $\bm{\beta}=(\beta_0, \beta_1)$ is updated in each iteration with starting value $m_0=(0,1)$ and the second hyperparameter is selected as $B=I_{2\times2}$. 
%The $\bm{\beta}$ is sensitive to the hyperparameters, especially the variance term $B$. 
This updating procedure for $m_0$ can help mitigate the sensitivity of hyperparameters.
%but is still restricted if the starting value is so far away from the true values. 
Regarding $B$, we recommend starting with the identity matrix and considering increasing or decreasing the diagonal items if the algorithm fails. For smoothing spline estimation, we set $k=15$. 
% The hyperparameters for $\bm{\beta}=(\beta_0, \beta_1)$ are $m0=(0,1)$ and  $B=\bigl(\begin{smallmatrix}
% 1&-0.8 \\ -0.8&1
% \end{smallmatrix} \bigr)$ for $n=500$ ,and $B=\bigl(\begin{smallmatrix}
% 1&0 \\ 0&1
% \end{smallmatrix} \bigr)$ for $n=100$. 
%The $\bm{\beta}$ is sensitive to the hyperparameters, especially the variance term $B$. We recommend starting with the identity matrix and considering adding a covariance term if the algorithm fails.

\begin{figure}[!tb]
    \centering
    \includegraphics[width=0.75\textwidth]{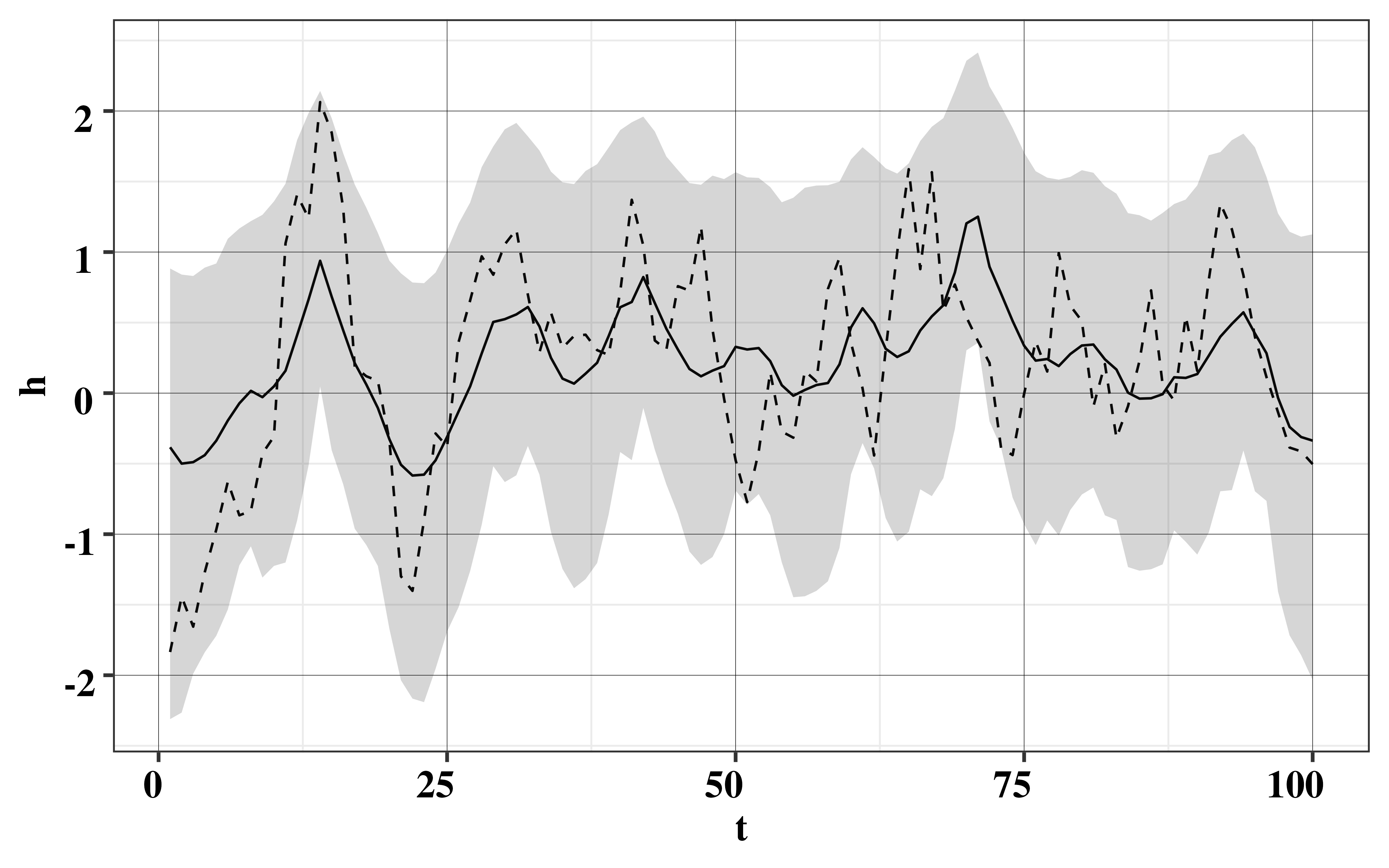}
    \caption{The true volatility $h$ (dashed line), posterior median volatility (solid line), and corresponding 95\% credible interval (shaded area) with sample size $n=100$ and $\beta_1=\ln(3)$ under the linear missing-data mechanism.}
    \label{fig:sim_post}
\end{figure}

\begin{figure}[!tb]
    \centering
    \includegraphics[width=0.75\textwidth]{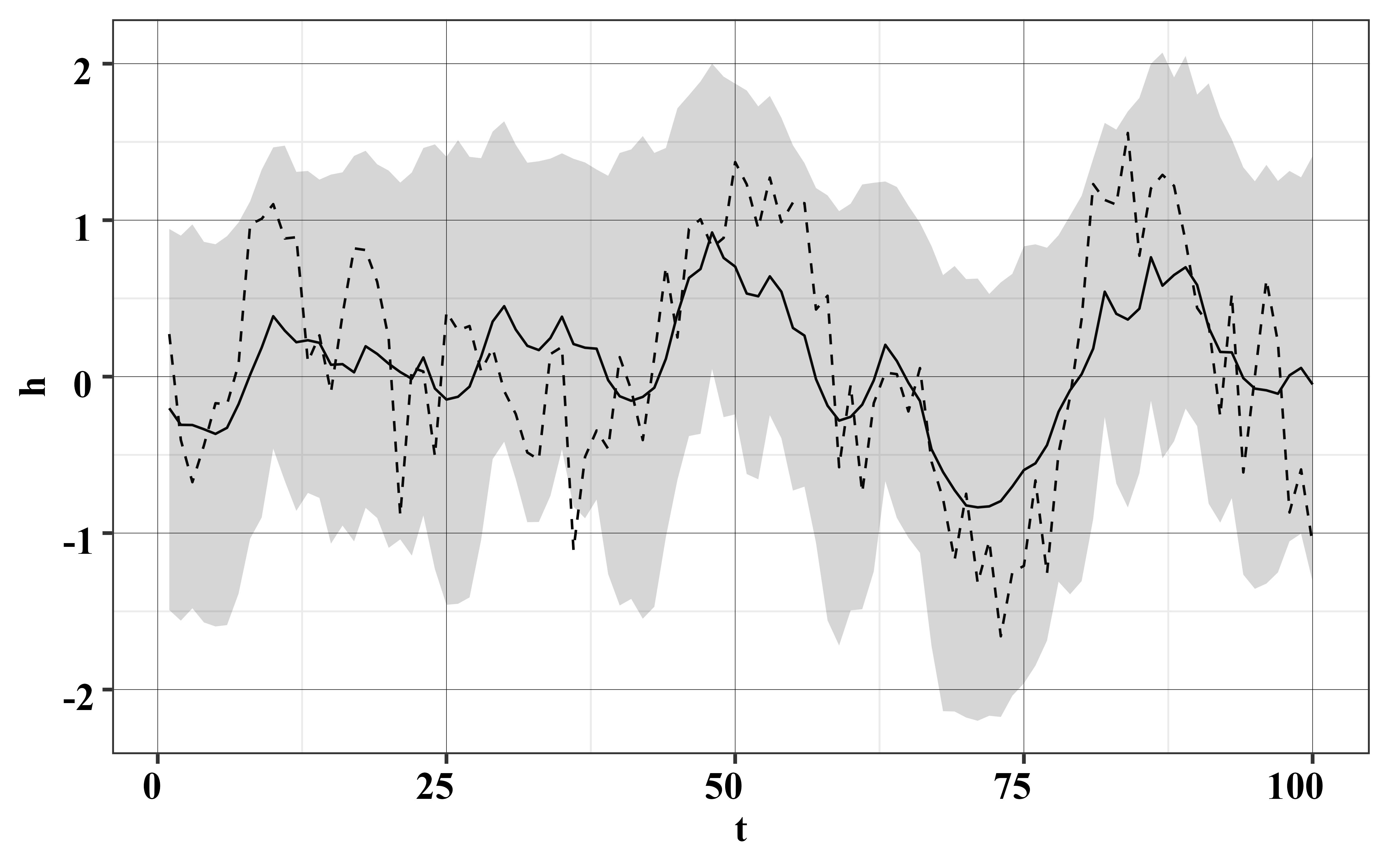}
    \caption{The true volatility $h$ (dashed line), posterior median volatility (solid line), and corresponding 95\% credible interval (shaded area) with sample size $n=100$ and smoothing spline missing-data mechanism with $\beta_1=\ln(4.5)$.}
    \label{fig:sim_post_ss}
\end{figure}

For all simulation settings, the number of particles is $N=20$. We run 500 replicates for each $\beta$ and sample size combination, and we run the Gibbs sampler for $32500$ iterations and discard the first $2500$ iterations as burn-in. For the linear missing-data mechanism, we set the initial value for the parameters as $(\mu^{[0]}, \phi^{[0]}, \sigma^{[0]}, \beta_0^{[0]}, \beta_1^{[0]})=\bigl(0.15, 0.9, \allowbreak 0.2, -1, 1\bigr)$.  Figure \ref{fig:sim_post} presents the true volatility $h$, posterior median volatility, and corresponding 95\% credible interval with sample size $n=100$ and $\beta_1=\ln(3)$ for a simulated series under the linear missing-data mechanism. For the smoothing spline missing-data mechanism, we set the initial value for the parameters as $(\mu^{[0]}, \phi^{[0]}, \sigma^{[0]})=\bigl(0.15, 0.9, \allowbreak 0.2\bigr)$, $\bm{d}^{[0]}=(0,-1)^{'}$, $\bm{c}^{[0]}=\bm{0}_{k\times 1} $, and $\lambda^{[0]}=\exp(-7)$. Figure \ref{fig:sim_post_ss} presents the true volatility $h$, posterior median volatility, and corresponding 95\% credible interval for a simulated series with sample size $n=100$ and $\beta_1=\ln(4.5)$.

% \begin{itemize}
%     \item Sampling $\phi$ and $\sigma$ together
% \end{itemize}

% We plotted the posterior coverage rates of $95\%$ credible interval for stochastic volatility($h_{1:n}$) in Figure \ref{fig:coverp500} and \ref{fig:coverp100}. 

% \begin{figure}[!htb]
%     \centering
%     \includegraphics[width=0.75\textwidth]{coverplot_n500.png}
%     \caption{Coverage rates of posterior credible interval for volatility(n=500)}
%     \label{fig:coverp500}
% \end{figure}

% \begin{figure}[!htb]
%     \centering
%     \includegraphics[width=0.75\textwidth]{coverp_n100.png}
%     \caption{Coverage rates of posterior credible interval for volatility(n=100)}
%     \label{fig:coverp100}
% \end{figure}

We calculated the mean posterior coverage rate and width of $95\%$ credible intervals over time, along with the average mean squared error (AMSE) over time for the stochastic volatility $h_{1:n}$. The summarized results are presented in Table \ref{tab:posth_both}. According to the results, the mean coverage rates show comparability across the two sample size groups. Within the same sample size category, the mean coverage rate is lower in the group with a larger odds ratio $\exp(\beta_1)$ of missingness under the logistic missing-data mechanism. The average mean squared error is smaller when the sample size is larger, but increases when the odds ratio of missingness is larger. 
%When sampling $\phi$ and $\sigma$ jointly, the mean coverage rates show comparability across the two sample size groups. Within the same sample size category, the mean coverage rate is lower in the group with a larger odds ratio ($\exp(\beta_1)$) of missingness. 
%, which is a result of a wider credible interval when the odds ratio of missingness is larger. 
% The average mean squared error is smaller in larger sample size groups but increases when the odds ratio of missingness is larger. 
%However, for sampling $\phi$ and $\sigma$ individually, the mean coverage rates for volatility are low in the small sample size group. Therefore, it is recommended to sample $\phi$ and $\sigma$ jointly when the sample size is small. 

\begin{table}[!htb]
\caption{Mean posterior coverage rate and width of $95\%$ credible interval over time and the average mean squared error(AMSE) over time for the stochastic volatility $h$  \label{tab:posth_both}}
\begin{center}
\begin{threeparttable}
\begin{tabular*}{\textwidth}{@{\extracolsep{\fill}}p{2.6cm} ccccc}
\cline{1-6}
Missing-data Mechanism & Sample size & $\exp(\beta_1)$\tnote{1} & AMSE & Mean width & Mean coverage rate\\
\cline{1-6}
%\hline
 \multirow{6}{4em}{Linear}&\multirow{3}{4em}{500} & 2.5 & 0.7598 & 2.3018 & 0.9290 \\
 & & 3 & 0.7962 & 2.3004	& 0.9185 \\
 & & 3.5 & 0.8055 & 2.2526	& 0.9054 \\
 \cline{2-6}
 & \multirow{3}{4em}{100} & 2.5 & 0.7811 & 2.3676 & 0.9359 \\
 & & 3 & 0.8130 & 2.3746	& 0.9307 \\
 & & 3.5 & 0.8400 & 2.3772	& 0.9250 \\
\cline{1-6}
%\hline
 \multirow{6}{4em}{Smoothing Spline}&\multirow{3}{4em}{500} & 2.5 & 0.7368 & 2.3411 & 0.9421 \\
& & 3.5 & 0.7403 & 2.3552	& 0.9442 \\
& & 4.5 & 0.7493 & 2.3763	& 0.9457 \\
 \cline{2-6}
&\multirow{3}{4em}{100} & 2.5 & 0.7753 & 2.4233 & 0.9495 \\
& & 3.5 & 0.7756 & 2.4372	& 0.9505 \\
& & 4.5 & 0.7847 & 2.4469	& 0.9503 \\
\cline{1-6}
%\hline
\end{tabular*}
%\begin{tablenotes}
%    \item [1] $\exp(\beta_1)$ is the odds ratio of missingness under the linear missing mechanism and is the parameter varying under the smoothing spline missing mechanism \eqref{ss_trueeq}
%\end{tablenotes}
\end{threeparttable}
\end{center}
\end{table}

We also conduct a comparison between the proposed method (method P) and two commonly used imputation techniques: imputing missing $y$ values with the mean of observed data (method A) and using the last observed data point for imputation (method B). We employ the same simulation settings. The summarized results for the stochastic volatility $h$ are presented in Table \ref{tab:mse_three}. The findings clearly indicate that the proposed method outperforms the other two imputation methods concerning the estimation of volatility series, and shows comparable performance in terms of model parameter estimation.% and sample $\phi$ and $\sigma$ jointly for all methods

\begin{table}[!htb]
\caption{Average mean squared error (AMSE) over time for the stochastic volatility $h$ \label{tab:mse_three}}
\begin{center}
%\centering
\begin{threeparttable}
\begin{tabular*}{\textwidth}{@{\extracolsep{\fill}}p{2.5cm}ccccc}
\cline{1-6}
& & & \multicolumn{3}{c}{AMSE}\\
 \cline{4-6}
Missing-data Mechanism &  Sample size & $\exp(\beta_1)$ & Method P & Method A & Method B\\%\tnote{1}
\cline{1-6}
\multirow{6}{4em}{Linear}& \multirow{3}{4em}{500} & 2.5 & 0.7598 & 0.9133 & 0.7891 \\
& & 3 & 0.7836 & 1.1203 & 0.8352 \\
& & 3.5 & 0.8055 & 1.4925 & 0.8906 \\
 \cline{2-6}
 & \multirow{3}{4em}{100} & 2.5 & 0.7811 & 0.8738 & 0.7987 \\
& & 3 & 0.8130 & 1.0103 & 0.8535 \\
& & 3.5 & 0.8400 & 1.1687 & 0.8981 \\
\cline{1-6}
 \multirow{6}{4em}{Smoothing Spline}& \multirow{3}{4em}{500} & 2.5 & 0.7368 & 0.8005 & 0.7786 \\
& & 3.5 & 0.7403 & 0.8229 & 0.7913 \\
& & 4.5 & 0.7493 & 0.8394 & 0.7961 \\
 \cline{2-6}
 & \multirow{3}{4em}{100} & 2.5 & 0.7753 & 0.8183 & 0.7825 \\
& & 3.5 & 0.7756 & 0.8219 & 0.7856 \\
& & 4.5 & 0.7847 & 0.9061 & 0.8725 \\
\cline{1-6}
\end{tabular*}
%\begin{tablenotes}
%    \item [1] $\exp(\beta_1)$ is the odds ratio of missingness under the linear missing mechanism and is the parameter varying under the smoothing spline missing mechanism \eqref{ss_trueeq}
%\end{tablenotes}
\end{threeparttable}
\end{center}
\end{table}

Additional results, including the mean posterior coverage rate, the width of the $95\%$ credible interval over time, and the mean squared error (AMSE) over time for model parameters, are provided in the supplementary materials.

\section{Application}
\label{application}
Suicide rates in the United States among young adults aged 18-24 have been consistently rising over the past two decades \citep{garnett2023}; there is an urgent need to identify ways to understand who is at the highest risk for suicide in the near future. Technological advancements in the form of mobile devices have given clinicians and researchers the ability to monitor those at risk at fine time scales.  These data can potentially be used to identify dynamic, proximal risk factors, which could be used to deploy time-sensitive interventions.  Our motivating application considers data collected at the University of Pittsburgh as part of a recently completed study to identify near-term risk factors for suicidal thoughts and behavior in young adults who had suicidal ideation and/or behavior in the past 4 months.  The data considered, which are displayed in Figure \ref{fig:happyori}, are EMA self-response from two study participants to the question ``At this moment, I feel happy'' using a visual analog scale slider with a scale of 0 to 100, where 0 indicates not all and 100 indicates very much.  The participants were prompted 7 times per day for 21 days, for a potential total of 147 observed responses. The mean (standard deviation) of the happiness responses are 29.43 (21.98) and 33.99 (27.92) for subjects one and two, respectively, and there $1-p = 4.1\%$ and $1-p = 8.8\%$ missing data where the participant did not respond to the prompt for subjects 1 and 2, respectively. %The mean and standard deviation of the happiness responses are 29.43 and 21.98, respectively, and there was a $1-p = 4.1\%$ rate of missing observations where the participant did not respond to the prompt. 

We analyzed the motivating data using the proposed procedure under both the simple logistic-linear missing-data mechanism and the cubic smoothing spline mechanism. Since the findings remain consistent across these two distinct missing-data mechanisms, we provide an elaborate discussion solely on the results obtained using the simple logistic-linear missing-data mechanism (Results pertaining to the cubic smoothing spline mechanism are available in the supplemental materials). The estimated posterior median (standard deviation) for $\beta_1$ was -0.059 (0.057) for subject 1 and -0.081 (0.051) for subject 2. This suggests that the odds of the participant not responding to the prompt are slightly higher for smaller values of happiness, which is consistent with disengagement during periods of low positive affect. The estimated posterior median of stochastic volatilities is displayed in Figure  \ref{fig:happypost}. We see a dynamic pattern in volatility, in particular a nadir between 10 and 16 days from the start of observation for subject 1 and between 3 and 7 days for subject 2.

\begin{figure}[!htb]%[!htb]
    \centering
    \includegraphics[width=.8\textwidth]{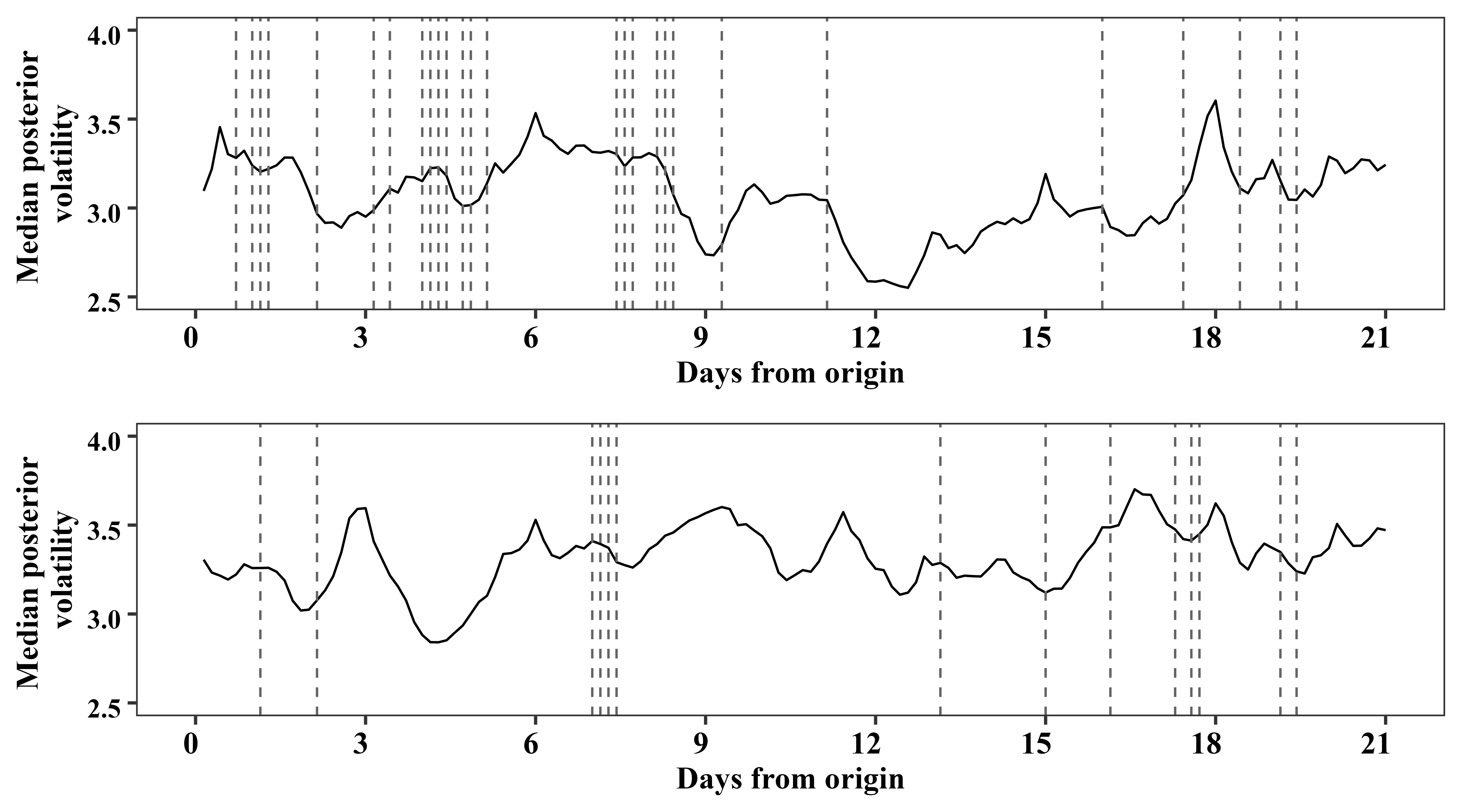}
    \caption{Estimated median posterior volatility of self-reported happiness (curve) along with time points when the participant reported that they wished to be dead (dotted vertical lines).}
    \label{fig:happypost}
\end{figure}

Researchers have found variability in mood to be associated with suicidal ideation \citep{claus2012}, where variability was quantified in two ways: as the mean squared successive difference between successive responses and as sample variance across the entire observation period.  Both of these methods inherently assume homogeneous variability across time. To the best of our knowledge, our current analysis is the first examination of the stochastic volatility of EMA in the context of suicidal risk.  To investigate the potential clinical associations and benefits of the fine-resolution time-course measure of variability provided by stochastic volatility, we consider its association with instances where the participant reported that they wished to be dead, which are indicated in Figure \ref{fig:happypost} through dotted vertical lines. We estimated time varying intensity function of events of wishing to be dead by applying Gaussian kernels to each data point and aggregating their contributions using the R function \texttt{density()} from the \texttt{stats} package. For kernel bandwidth selection, we utilized the default rule of thumb method. 
% regression spline of the log-intensity of a nonhomogenous Poisson process 
% using the R package \textit{fda} \citep{fdapack}
We then computed Pearson's correlation between this intensity and estimated volatility across time for each subject. The estimated correlation between volatility of happiness and intensity of wishing to be dead is 0.4518 with a p-value less than 0.0001 for subject 1, and 0.3769 with a p-value less than 0.0001 for subject 2. %The estimated correlation between volatility of happiness and intensity of wishing to be dead is 0.4737 with a p-value less than 0.0001.  

As self-reported EMA levels of happiness themselves and their means have been shown to be associated with suicide ideation \citep{kivela2022}, it is important to compare the strengths of association between suicide ideation and volatility of happiness to its associations with reported levels of happiness and mean happiness. The estimated correlation between the intensity of the participant wishing that they were dead and self-reported happiness is 0.0377 with a p-value of 0.6570 for subject 1 and -0.0271 with a p-value of 0.7562 for subject 2.
%-0.0778 with a p-value of 0.3592. 
In addition, we applied a generalized cross-validated smoothing spline estimator to the self-reported time series of happiness to obtain a de-noised time-varying estimate of mean happiness.  The correlation between this smoothed estimate of mean happiness and the intensity of a participant wishing that they were dead was estimated as 0.0699 with a p-value of 0.4001 for subject 1 and -0.0049 with a p-value of 0.9528 for subject 2.
%-0.0563 with a p-value of 0.4979.  
These results, where stochastic volatility has a stronger correlation with a measure of suicide ideation, suggest a potential use in improving monitoring and predicting suicide ideation beyond that provided by reported levels of mood.  

\begin{comment}
\begin{table}[!htb]
\caption{Correlation and corresponding p-values for correlation test between the intensity function and estimated volatility($h$), original mood series($y$), and spline-smoothed mood series($\tilde{y}$). \label{tab:cortest_happy}}
\begin{center}
\begin{tabular*}{.53\textwidth}{ccc}
\hline
Measure & Pearson's correlation & p-value\\\hline
 $h$ & 0.3088 & 0.0001\\
 $y$ & 0.2122	& 0.0115 \\
 $\tilde{y}$ & 0.2075	& 0.0117 \\
\hline
\end{tabular*}
\end{center}
\end{table}
\end{comment}

\section{Discussion}
\label{discussion}

In this paper, we introduced the first method for fitting stochastic volatility models with informed missingness.  The approach is built on an imputation algorithm for the stochastic volatility model with informative missingness based on Tukey's representation and extended the conditional particle filter with ancestor sampling algorithm to account for the imputation.  Although we focused on the stochastic volatility model in this paper, the conditional particle filter with ancestor sampling algorithm that is the foundation of the proposed method is developed for nonlinear state space models \citep{Lindsten15}. Thus, the proposed method can potentially be extended to other nonlinear state space models by modifying the imputation step according to the model settings.  

%From the simulation and application results, the proposed method works well in estimating stochastic volatility. The application results also illustrate the association between the estimated volatility and outcome of interest, which is consistent with our motivation and previous researches\citep{Renee11, Kuppens, McConville, Angst}. 

%This work can be improved by modifying the sampling method for $\beta_0$ and $\beta_1$ so that the results are less sensitive to the choice of hyperparameters in the prior distribution. Extending the work from univariate to multivariate could also be of interest, as this could incorporate more information into the model, such as the relationship between mood measures. Furthermore, we are planning to develop a method for the prediction of the outcome of interest using the estimated volatility series in future work, which could potentially be utilized in warning systems for suicide studies. 

This presents the first approach to the analysis of stochastic volatility with informed missingness and considers the simplest case of single realizations of univariate conditionally Gaussian data.  It leaves many open questions that can be the focus of future work.  The first line of extensions is with regards to the form of responses, including methods for the analysis of multivariate time series and for ordinal time series, which are common forms of EMA responses. Although methods for defining multivariate stochastic volatility models exist, primarily via factor models, how to account for different patterns of missingness among different variables could be challenging.  A second line of important extensions is to second-level analyses from multiple subjects.  This article considered the estimation from data from a single subject, and used exploratory analyses to evaluate the potential use of volatility as a clinical measure.  Future work can extend the single subject model into a multilevel model for multiple subjects, as well as joint models for the simultaneous modeling of stochastic volatility and prediction of time to events.

\bibliographystyle{agsm}
\bibliography{refs}

\begin{appendices}

\section{Imputation procedure for informative missingess}\label{appendix:imp}

Following the notations introduced in Sections 2 and 4, we provide supplemental details on the derivation of the proposed imputation procedure.

\subsection{Details of the Tukey's representation}\label{appdenix:tukeyrep}

First note that
\begin{equation*}
    f\left( y_t|\theta_{Y|R}, h_{t-1}\right)=pf(y_t|r_t=1,\theta_{Y|R}, h_{t-1})+(1-p)f(y_t|r_t=0,\theta_{Y|R}, h_{t-1})
    %f()=\frac{p}{1-p}\frac{1-r(y_t)}{r(y_t)}f(y_t|r_t=1,\theta_{Y|R}, h_{t-1})
\end{equation*}
Thus,
\begin{equation*}
    r(y_t)=P(r_t=1|y_t,\Theta, h_{t-1})=\frac{pf(y_t|r_t=1,\theta_{Y|R}, h_{t-1})}{f\left( y_t|\theta_{Y|R}, h_{t-1}\right)}
\end{equation*}
The logit of $r(y_t)$ is given by
\begin{align}
    \logit\left(\frac{r(y_t)}{1-r(y_t)}\right)& =\log\left[ r(y_t)\right]-\log\left[ 1-r(y_t)\right] \nonumber\\
    &= \log\left[ \frac{pf(y_t|r_t=1,\theta_{Y|R}, h_{t-1})}{f\left( y_t|\theta_{Y|R}, h_{t-1}\right)}\right]-\log\left[ \frac{(1-p)f(y_t|r_t=0,\theta_{Y|R}, h_{t-1})}{f\left( y_t|\theta_{Y|R}, h_{t-1}\right)}\right] \nonumber\\
    &= \log\left[ \frac{p}{1-p}\right]+\log\left[ f(y_t|r_t=1,\theta_{Y|R}, h_{t-1})\right]
    \nonumber\\
    &-\log\left[ f(y_t|r_t=0,\theta_{Y|R}, h_{t-1})\right]\label{appendix:tukey0}
\end{align}
The equation \eqref{appendix:tukey0} can be rewritten as 
\begin{equation}
   f(y_t|r_t=0,\theta_{Y|R}, h_{t-1})=\frac{p}{1-p}\frac{1-r(y_t)}{r(y_t)}f(y_t|r_t=1,\theta_{Y|R}, h_{t-1}).
   \label{appendix:tukey}
\end{equation}

\subsection{Details of the implied distribution for missing data $y_t$}\label{appdenix:impf}

As introduced in Section 4 of the main article, under the univariate stochastic volatility model setting, we have

\begin{align}
    \frac{1-r(y_t)}{r(y_t)} & =\frac{1}{exp(\beta_0+\beta_1 y_t)}\\
    \frac{p}{1-p} & =\exp \left(\beta_0-\frac{\beta_1\sigma_t^2}{2}\right)\\
    y_t \mid & r_t=1,\Theta, h_{t-1}  \sim N(0,\sigma_t^2).
\end{align}

Applying Tukey's representation \ref{appendix:tukey}, the implied distribution for missing $y_t$ can be derived as

\begin{align*}
    f(y_t|r_t=0,\theta_{Y|R}, h_{t-1}) & =\frac{p}{1-p}\frac{1-r(y_t)}{r(y_t)}f(y_t|r_t=1,\theta_{Y|R}, h_{t-1})\\
    & = \exp \left(\beta_0-\frac{\beta_1\sigma_t^2}{2}\right) \frac{1}{exp(\beta_0+\beta_1 y_t)} \times \frac{1}{\sqrt{2\pi \sigma_t^2}}exp(\frac{-y_t^2}{2\sigma_t^2})\\
    %& = \frac{p}{1-p} exp(-\beta_0+\frac{\beta_1\sigma_t^2}{2}) \frac{1}{\sqrt{2\pi \sigma_t^2}}exp(\frac{(y_t+\beta_1 \sigma_t^2)^2}{2\sigma_t^2})  
    & = \frac{1}{\sqrt{2\pi \sigma_t^2}}exp(\frac{-(y_t+\beta_1 \sigma_t^2)^2}{2\sigma_t^2})\\ 
    & \sim N(-\beta_1\sigma_t^2,\sigma_t^2).
\end{align*}

\section{Additional details of the Gibbs sampler}\label{appdenix:gibbspara}

For a more comprehensive understanding of the Gibbs sampler, as originally introduced in Section 6 of the main article, we present additional details on sampling $\sigma$ and $\phi$ in this section.

%\begin{enumerate}
    %\item[(i)] Details on sampling $\sigma$ and $\phi$

    We provide details for two priors and approaches for sampling $\sigma$ and $\phi$. The first approach assumes independent priors where  
    We used two methods for sampling $\phi, \sigma^2$. The first method involves sampling $\phi$ and $\sigma^2$ separately.
    \begin{enumerate}
        \item [(a)] $\phi$

            We used a method based on the Metropolis-Hastings algorithm\citep{CHIB94, KimS98}. 
            \begin{itemize}
                \item prior: 
                $$\frac{1+\phi}{2} \sim Beta(a,b)$$
                $$\pi(\phi)\propto \{\frac{1+\phi}{2}\}^{a-1}\{\frac{1-\phi}{2}\}^{b-1}$$
                where $a$, $b$ are hyperparameters that satisfy $a,b >\frac{1}{2}$.
                \item posterior and acceptance sampling method:
                At $i$th iteration, given $\phi^{(i-1)}$ from $i-1$th iteration, sample $\phi^{*}$ from proposal distribution $N(\hat{\phi},V_{\phi})$, where $\hat{\phi}=\frac{\sum_{t=1}^{n-1}(h_{t+1}-\mu)(h_t-\mu)}{\sum_{t=1}^{n-1}(h_t-\mu)^2}$ and $V_{\phi}=\frac{\sigma^2}{\sum_{t=1}^{n-1}(h_t-\mu)^2}$. Then if the proposal value is in the stationary region, it will be accepted with probability $exp(g(\phi^{*})-g(\phi^{(i-1)})$, where
                $$g(\phi)=log\pi(\phi)-\frac{(h_1-\mu)^2(1-\phi^2)}{2\sigma^2}+\frac{1}{2}log(1-\phi^2).$$ If $\phi^{*}$ is rejected, set $\phi^{(i)}$ as $\phi^{(i-1)}.$
            \end{itemize}
        \item[(b)] $\sigma^2$

            \begin{itemize}
                \item prior: A Half-t prior distribution \citep{gelman06,zeda21} is used for $\sigma$
                $$\sigma \sim \text{Half-}t(v,G)$$
                $$\pi(\sigma)\propto \{1+\frac{1}{v}(\frac{\sigma}{G})^2\}^{-(v+1)/2}, \sigma>0$$
                where $v$, $G$ are hyperparameters.
                This distribution is equivalent to a scale mixture of inverse Gamma distribution(IG) for $\sigma^2$ \citep{wand11}: 
                $$\sigma^2|q_{\sigma}\sim IG(\frac{v}{2},\frac{v}{q_{\sigma}}),$$
                $$q_{\sigma} \sim IG(\frac{1}{2},\frac{1}{G^2}).$$
                \item posterior:
                $$\sigma^2|q_{\sigma},h_{1:n},y_{1:n},\phi,\mu \sim IG(\frac{v+n}{2},A)$$
                $$q_{\sigma}|\sigma^2,h_{1:n},y_{1:n},\phi,\mu \sim IG(\frac{v+1}{2},\frac{1}{G^2}+\frac{v}{\sigma^2})$$
                where
                $$A=\frac{v}{q_{\sigma}}+\frac{1}{2}\{(1-\phi^2)(h_1-\mu)^2+\sum_{t=2}^n [(h_t-\mu)-\phi(h_{t-1}-\mu)^2]\}$$
            \end{itemize}
    \end{enumerate}

Another method involves sampling $\phi$ and $\sigma^2$ together from posterior distribution $f(\phi,\sigma|\mu,y_{1:T},h_{1:T})$ using Random Walk Metropolis-Hastings(RWMH) algorithm \citep{Chen21}. The prior and posterior for $\left( \phi,\sigma \right)$ are presented below and the Random Walk Metropolis-Hastings(RWMH) algorithm is summarized in the Algorithm \ref{alg:rwmh}.

\begin{itemize}
    \item prior:
    $$\binom{\phi}{\sigma}\sim N(\binom{\mu_{\phi}}{\mu_{\sigma}},\Sigma)$$
    where
    \[ \Sigma = 
                \begin{bmatrix}
                \sigma_{\phi}^2 & \rho \sigma_{\phi} \sigma_{q}\\
                \rho \sigma_{\phi} \sigma_{q} & \sigma_{q}^2
           \end{bmatrix}
           \]
    \item posterior:
    $$f(\phi,\sigma|\mu,y_{1:T},h_{1:T}) \propto \pi(\phi,\sigma)f(h_1|\phi,\sigma,\mu)\prod_{i=2}^T{f(h_i|h_{i-1},\phi,\sigma,\mu)}$$
\end{itemize}

\begin{algorithm}[H]
\caption{Random Walk Metropolis-Hastings(RWMH)}
\label{alg:rwmh}
%\hspace*{\algorithmicindent}  \\
\begin{algorithmic}[1]
    \STATE The initial state $\xi^0=(\phi,\sigma)$.
    \FOR{$r\geq0$}
        \STATE Draw $\epsilon_r\sim N(0,\lambda \Sigma_{\epsilon})$ and let $\eta^r=\xi^{r-1}+\epsilon_r$.
        \STATE Compute the acceptance probability $\alpha=min(\frac{g(\xi)}{g(\xi^r)},1)$.
        \STATE Generate $U_r\sim unif(0,1)$.
        \IF{$U_r<\alpha$}
            \STATE Set $\xi^r=\eta^r$
        \ELSIF{$U_r\geq \alpha$}
            \STATE Set $\xi^r=\xi^{r-1}$
        \ENDIF
    \ENDFOR
\end{algorithmic}
\end{algorithm}

where
\begin{align*}   
    g(\xi) &= exp\{ -\frac{(\phi-\mu_{\phi})^2\sigma_q^2+(\sigma-\mu_q)^2\sigma_{\phi}^2-2\rho \sigma_{\phi} \sigma_q(\phi-\mu_{\phi})(\sigma_-\mu_q)}{2(1-\rho^2)\sigma_{\phi}^2\sigma_q^2} \} \\
    & \times \frac{\sqrt{1-\phi^2}}{\sigma^n} exp\{ -\frac{(1-\phi^2)(h_1-\mu)^2+\sum_{i=2}^n[(h_i-\mu)-\phi(h_{i-1}-\mu)]^2}{2\sigma^2} \}
\end{align*}

\section{Additional results for simulation studies}\label{appdenix:simul}

In Section 6 of the main manuscript, we present the posterior results for the stochastic volatility series. In this section, we provide the mean posterior coverage rate and width of $95\%$ credible intervals and mean squared error for model parameters $\Theta =\left( \sigma, \phi, \mu, \beta_0, \beta_1 \right)$.
%In Section 7 of the main manuscript, we present the posterior results for the stochastic volatility. We provide the mean posterior coverage rate and width of $95\%$ credible intervals and mean squared error for model parameters $\Theta =\left( \sigma, \phi, \mu, \beta_0, \beta_1 \right)$. According to the results, sampling $\phi$ and $\sigma$ jointly is outperformed over sampling $\phi$ and $\sigma$ individually when the sample size is small (100).

\begin{table}[!htb]
\caption{Mean posterior coverage rate and width of $95\%$ credible intervals and mean squared error for model parameters under linear missing mechanism\label{tab:postpara}}
\begin{center}
\begin{tabular*}{\textwidth}{p{2cm} ccccc}
\hline
Parameter & Sample size & $\exp(\beta_1)$ & Mean coverage rate & Mean width & MSE \\\hline
 \multirow{6}{4em}{$\sigma$}&\multirow{3}{4em}{500} & 2.5 & 0.9580 & 0.2477&0.0076 \\
 & & 3.5 & 0.9900&0.2643&0.0069 \\
 & & 4.5 & 0.9020&0.2428&0.0095 \\
 \cline{2-6}
 & \multirow{3}{4em}{100} & 2.5 & 1.0000 & 0.3144& 0.0098 \\
 & & 3.5 & 0.9979 & 0.3116 & 0.0102 \\
 & & 4.5 & 0.9979 & 0.3123 & 0.0107 \\
\hline
 \multirow{6}{4em}{$\phi$}&\multirow{3}{4em}{500} & 2.5 & 0.9920 & 0.1829 & 0.0033 \\
 & & 3.5 & 0.9880 & 0.2263 & 0.0061 \\
 & & 4.5 & 0.9980 & 0.1931 & 0.0034 \\
 \cline{2-6}
 & \multirow{3}{4em}{100} & 2.5 & 0.9959 & 0.2285 & 0.0057 \\
 & & 3.5 & 1.0000 & 0.2320 & 0.0058 \\
 & & 4.5 & 0.9959 & 0.2351 & 0.0058 \\
\hline
 \multirow{6}{4em}{$\mu$}&\multirow{3}{4em}{500} & 2.5 & 0.9980 & 0.3995 & 0.0151 \\
 & & 3.5 & 0.9460 & 0.3283 & 0.0127 \\
 & & 4.5 & 0.9540 & 0.3502 & 0.0150 \\
 \cline{2-6}
 & \multirow{3}{4em}{100} & 2.5 & 1.0000 & 0.8928 & 0.0689 \\
 & & 3.5 & 1.0000 & 0.8787 & 0.0715 \\
 & & 4.5 & 1.0000 & 0.8757 & 0.0743 \\
\hline
 \multirow{6}{4em}{$\beta_1$}&\multirow{3}{4em}{500} & 2.5 & 0.8120 & 1.3196 & 0.4408 \\
 & & 3.5 & 0.8800 & 1.0254 & 0.1921 \\
 & & 4.5 & 0.8560 & 1.0475 & 0.2644 \\
 \cline{2-6}
 & \multirow{3}{4em}{100} & 2.5 & 0.8330 & 1.6966 & 0.5985 \\
 & & 3.5 & 0.8274 & 2.1267 & 0.8541 \\
 & & 4.5 & 0.8066 & 1.6917 & 0.6569 \\
\hline
\end{tabular*}
\end{center}
\end{table}

\begin{table}[!htb]
\caption{Mean posterior coverage rate and width of $95\%$ credible intervals and mean squared error for model parameters under cubic smoothing spline missing mechanism\label{tab:postpara_ss}}
\begin{center}
\begin{tabular*}{\textwidth}{p{2cm} ccccc}
\hline
Parameter & Sample size & $\exp(\beta_1)$ & Mean coverage rate & Mean width & MSE \\\hline
 \multirow{6}{4em}{$\sigma$}&\multirow{3}{4em}{500} & 2.5 & 0.9733 & 0.2489&0.0067 \\
 & & 3.5 & 0.9868&0.2533&0.0069 \\
 & & 4.5 & 0.9957&0.2568&0.0066 \\
 \cline{2-6}
 & \multirow{3}{4em}{100} & 2.5 & 1.0000 & 0.3216& 0.0091 \\
 & & 3.5 & 1.0000 & 0.3213 & 0.0090 \\
 & & 4.5 & 1.0000 & 0.3258 & 0.0090 \\
\hline
 \multirow{6}{4em}{$\phi$}&\multirow{3}{4em}{500} & 2.5 & 0.9756 & 0.1792 & 0.0033 \\
 & & 3.5 & 0.9868 & 0.1797 & 0.0033 \\
 & & 4.5 & 0.9762 & 0.1818 & 0.0033 \\
 \cline{2-6}
 & \multirow{3}{4em}{100} & 2.5 & 0.9931 & 0.2290 & 0.0057 \\
 & & 3.5 & 0.9911 & 0.2288 & 0.0057 \\
 & & 4.5 & 0.9977 & 0.2314 & 0.0055 \\
\hline
 \multirow{6}{4em}{$\mu$}&\multirow{3}{4em}{500} & 2.5 & 1.0000 & 0.4756 & 0.0200 \\
 & & 3.5 & 1.0000 & 0.4793 & 0.0199 \\
 & & 4.5 & 1.0000 & 0.4897 & 0.0203 \\
 \cline{2-6}
 & \multirow{3}{4em}{100} & 2.5 & 1.0000 & 0.9629 & 0.1021 \\
 & & 3.5 & 1.0000 & 0.9717 & 0.0796 \\
 & & 4.5 & 1.0000 & 0.9705 & 0.0836 \\
\hline
\end{tabular*}
\end{center}
\end{table}

The posterior results for parameters for the proposed method and commonly used ad hoc methods are also provided in this section. The AMSEs for model parameters using the proposed method are comparable to those from ad hoc methods.

\begin{table}[h]%!htb
\caption{Average mean squared error(AMSE) over time for the stochastic volatility $h$ and model parameters under linear missing mechanism\label{tab:mse_three}}
\begin{center}
%\centering
\begin{tabular*}{0.85\textwidth}{cccccc}
\cline{1-6}
& & & \multicolumn{3}{c}{AMSE}\\
 \cline{4-6}
Parameter & Sample size & $\exp(\beta_1)$ & Method P & Method A & Method B\\
\cline{1-6}
\multirow{6}{4em}{h}& \multirow{3}{4em}{500} & 2.5 & 0.7598 & 0.9133 & 0.7891 \\
& & 3 & 0.7836 & 1.1203 & 0.8352 \\
& & 3.5 & 0.8055 & 1.4925 & 0.8906 \\
 \cline{2-6}
 & \multirow{3}{4em}{100} & 2.5 & 0.7811 & 0.8738 & 0.7987 \\
& & 3 & 0.8130 & 1.0103 & 0.8535 \\
& & 3.5 & 0.8400 & 1.1687 & 0.8981 \\
\cline{1-6}
 \multirow{6}{4em}{$\sigma$}& \multirow{3}{4em}{500} & 2.5 & 0.0076	& 0.0133 & 0.0069 \\
& & 3 & 0.0069 & 0.0261 & 0.0067 \\
& & 3.5 & 0.0095 & 0.0742 & 0.0071 \\
 \cline{2-6}
& \multirow{3}{4em}{100} & 2.5 & 0.0098 & 0.0098 & 0.0095 \\
& & 3 & 0.0102 & 0.0107 & 0.0096 \\
& & 3.5 & 0.0107 & 0.0117 & 0.0098 \\
\cline{1-6}
 \multirow{6}{4em}{$\phi$}& \multirow{3}{4em}{500} & 2.5 & 0.0033 & 0.0040 & 0.0032 \\
& & 3 & 0.0061 & 0.0057 & 0.0030 \\
& & 3.5 & 0.0034 & 0.0081 & 0.0032 \\
 \cline{2-6}
& \multirow{3}{4em}{100} & 2.5 & 0.0057 & 0.0051 & 0.0059 \\
& & 3 & 0.0058 & 0.0049 & 0.0058 \\
& & 3.5 & 0.0058 & 0.0049 & 0.0059 \\
\cline{1-6}
 \multirow{6}{4em}{$\mu$}& \multirow{3}{4em}{500} & 2.5 & 0.0151 & 0.0219 & 0.0160 \\
& & 3 & 0.0127 & 0.0431 & 0.0160 \\
& & 3.5 & 0.0150 & 0.0919 & 0.0176 \\
 \cline{2-6}
& \multirow{3}{4em}{100} & 2.5 & 0.0689 & 0.0718 & 0.0685 \\
& & 3 & 0.0715 & 0.0819 & 0.0775 \\
& & 3.5 & 0.0743 & 0.1116 & 0.0739 \\
\cline{1-6}
\end{tabular*}
\end{center}
\end{table}

\begin{table}[h]%
\caption{Average mean squared error(AMSE) over time for the stochastic volatility $h$ and model parameters under cubic smoothing spline missing mechanism\label{tab:mse_three_ss}}
\begin{center}
%\centering
\begin{tabular*}{0.85\textwidth}{cccccc}
\cline{1-6}
& & & \multicolumn{3}{c}{AMSE}\\
 \cline{4-6}
Parameter & Sample size & $\exp(\beta_1)$ & Method P & Method A & Method B\\
\cline{1-6}
\multirow{6}{4em}{h}& \multirow{3}{4em}{500} & 2.5 & 0.7368 & 0.8005 & 0.7786 \\
& & 3.5 & 0.7403 & 0.8229 & 0.7913 \\
& & 4.5 & 0.7493 & 0.8394 & 0.7961 \\
 \cline{2-6}
 & \multirow{3}{4em}{100} & 2.5 & 0.7753 & 0.8183 & 0.7825 \\
& & 3.5 & 0.7756 & 0.8219 & 0.7856 \\
& & 4.5 & 0.7847 & 0.9061 & 0.8725 \\
\cline{1-6}
 \multirow{6}{4em}{$\sigma$}& \multirow{3}{4em}{500} & 2.5 & 0.0067	& 0.0117 & 0.0082 \\
& & 3.5 & 0.0069 & 0.0146 & 0.0090 \\
& & 4.5 & 0.0066 & 0.0166 & 0.0091 \\
 \cline{2-6}
& \multirow{3}{4em}{100} & 2.5 & 0.0091 & 0.0109 & 0.0096 \\
& & 3.5 & 0.0090 & 0.0112 & 0.0094 \\
& & 4.5 & 0.0090 & 0.0172 & 0.0141 \\
\cline{1-6}
 \multirow{6}{4em}{$\phi$}& \multirow{3}{4em}{500} & 2.5 & 0.0033 & 0.0030 & 0.0030 \\
& & 3.5 & 0.0033 & 0.0029 & 0.0030 \\
& & 4.5 & 0.0033 & 0.0030 & 0.0030 \\
 \cline{2-6}
& \multirow{3}{4em}{100} & 2.5 & 0.0057 & 0.0055 & 0.0061 \\
& & 3.5 & 0.0057 & 0.0054 & 0.0061 \\
& & 4.5 & 0.0055 & 0.0051 & 0.0053 \\
\cline{1-6}
 \multirow{6}{4em}{$\mu$}& \multirow{3}{4em}{500} & 2.5 & 0.0200 & 0.0179 & 0.0194 \\
& & 3.5 & 0.0199 & 0.0185 & 0.0206 \\
& & 4.5 & 0.0203 & 0.0191 & 0.0189 \\
 \cline{2-6}
& \multirow{3}{4em}{100} & 2.5 & 0.1021 & 0.0855 & 0.1007 \\
& & 3.5 & 0.0796 & 0.0813 & 0.0963 \\
& & 4.5 & 0.0836 & 0.1151 & 0.0989 \\
\cline{1-6}
\end{tabular*}
\end{center}
\end{table}

\begin{table}[h]%!htb
\caption{Mean posterior coverage rate and width of $95\%$ credible interval over time and the average mean squared error(AMSE) over time for smoothing spline function  \label{tab:postell_ss}}
\begin{center}
\begin{tabular*}{0.85\textwidth}{ccccc}%
\hline
Sample size & $\exp(\beta_1)$ & AMSE & Mean width & Mean coverage rate\\\hline
\multirow{3}{4em}{500} & 2.5 & 19.8800 & 7.3030 & 0.8974 \\
& 3.5 & 20.4429 & 6.9856	& 0.8594 \\
& 4.5 & 22.5088 & 7.4068	& 0.8203 \\
\hline
\multirow{3}{4em}{100} & 2.5 & 26.5555 & 12.3529 & 0.9732 \\
& 3.5 & 27.0224 & 11.9002	& 0.9647 \\
& 4.5 & 27.6660 & 12.0874	& 0.9602 \\
\hline
\end{tabular*}
\end{center}
\end{table}

\section{Additional results for the application}\label{appdenix:application}

We analyzed the motivating data using the proposed procedure under the cubic smoothing spline mechanism and the findings are consistent with those under the logistic missing mechanism. 

\begin{figure}[!htb]%[!htb]
    \centering
    \includegraphics[width=.8\textwidth]{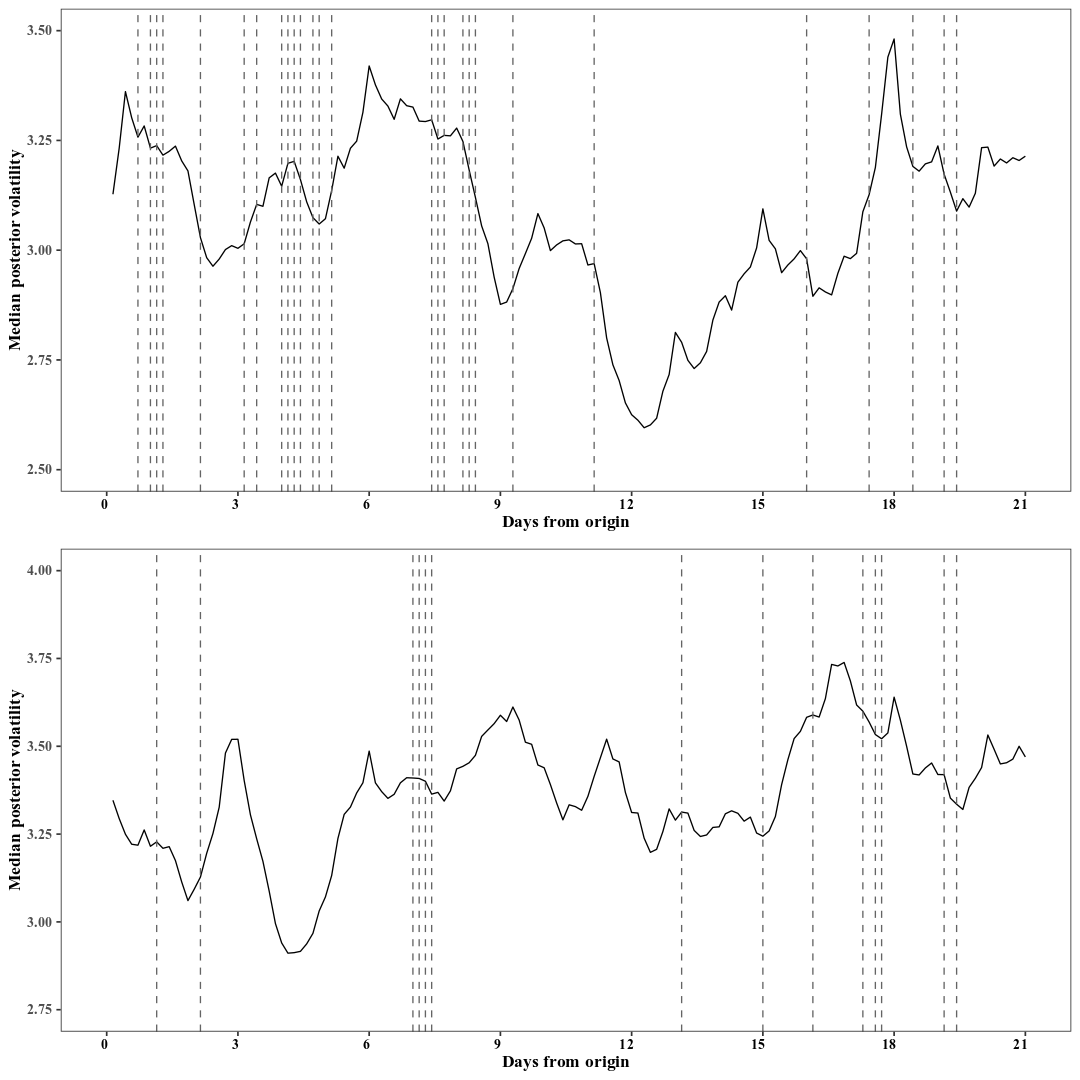}
    \caption{Estimated median posterior volatility of self-reported happiness (curve) along with time points when the participant reported that they wished to be dead (dotted vertical lines).}
    \label{fig:happypost}
\end{figure}

\begin{table}[!htb]
\caption{Correlation and corresponding p-values for correlation test between the intensity function and estimated volatility($h$), original mood series($y$), and spline-smoothed mood series($\tilde{y}$). \label{tab:cortest_happy}}
\begin{center}
\begin{tabular*}{.7\textwidth}{cccc}
\hline
    Subject&Measure & Pearson's correlation & p-value\\
    \hline
     \multirow{3}{2em}{1}&$h$ & $0.5257$ & $<0.0001$\\
     &$y$ & $0.0377$	& $0.6570$ \\
     &$\tilde{y}$ & $0.0699$	& $0.4001$ \\
    \hline
    \multirow{3}{2em}{2}&$h$ & $0.5252$ & $<0.0001$\\
     &$y$ & $-0.0271$	& $0.7562$ \\
     &$\tilde{y}$ & $-0.0049$	& $0.9528$ \\
     \hline
\end{tabular*}
\end{center}
\end{table}

\end{appendices}

\end{document}